# THE TWO SOURCES OF SOLAR ENERGETIC PARTICLES

DONALD V. REAMES

*Institute for Physical Science and Technology*
*University of Maryland, College Park, MD*
*Email: dvreames@umd.edu*

**Abstract**. Evidence for two different physical mechanisms for acceleration of solar energetic particles (SEPs) arose 50 years ago with radio observations of type III bursts, produced by outward streaming electrons, and type II bursts from coronal and interplanetary shock waves. Since that time we have found that the former are related to "impulsive" SEP events from impulsive flares or jets. Here, resonant stochastic acceleration, related to magnetic reconnection involving open field lines, produces not only electrons but 1000-fold enhancements of $^{3}$He/$^{4}$He and of (Z>50)/O. Alternatively, in "gradual" SEP events, shock waves, driven out from the Sun by coronal mass ejections (CMEs), more democratically sample ion abundances that are even used to measure the coronal abundances of the elements. Gradual events produce by far the highest SEP intensities near Earth. Sometimes residual impulsive suprathermal ions contribute to the seed population for shock acceleration, complicating the abundance picture, but this process has now been modeled theoretically. Initially, impulsive events define a point source on the Sun, selectively filling few magnetic flux tubes, while gradual events show extensive acceleration that can fill half of the inner heliosphere, beginning when the shock reaches ~2 solar radii. Shock acceleration occurs as ions are scattered back and forth across the shock by resonant Alfvén waves amplified by the accelerated protons themselves as they stream away. These waves also can produce a streaming-limited maximum SEP intensity and plateau region upstream of the shock. Behind the shock lies the large expanse of the "reservoir", a spatially extensive trapped volume of uniform SEP intensities with invariant energy-spectral shapes where overall intensities decrease with time as the enclosing "magnetic bottle" expands adiabatically. These reservoirs now explain the slow intensity decrease that defines gradual events and was once erroneously attributed solely to slow outward diffusion of the particles. At times the reservoir from one event can contribute its abundances and even its spectra as a seed population for acceleration by a second CME-driven shock wave. Confinement of particles to magnetic flux tubes that thread their source early in events is balanced at late times by slow velocity-dependent migration through a tangled network produced by field-line random walk that is probed by SEPs from both impulsive and gradual events and even by anomalous cosmic rays from the outer heliosphere. As a practical consequence, high-energy protons from gradual SEP events can be a significant radiation hazard to astronauts and equipment in space and to the passengers of high-altitude aircraft flying polar routes.

**Table of Contents**

## 1. Introduction

The identification and description of two physical mechanisms involved in the acceleration of solar energetic particles (SEPs) began long ago and has since been supported by a wide range of observations and theoretical studies. Before reviewing recent results and our modern understanding of the properties of the SEPs and their sources, we begin by tracing the history, and the origin and evolution of our current ideas.

The first SEP events, reported by Forbush (1946), were what we now call ground-level events (GLEs). They are produced when ~GeV protons create a nuclear cascade through the Earth's atmosphere that can be observed by detectors at ground level as an increase above the continuous background produced in a similar manner by galactic cosmic rays. Most of the 71 GLEs of the last 70 years barely exceed this background by a few percent, the largest by a factor of 45 (Cliver *et al.* 1982). Since solar flares had been observed by Carrington (1860) much earlier, it was natural to attribute SEP events to some (unspecified) process associated with solar flares. However, radio observations reviewed by Wild, Smerd, and Weiss (1963) suggested evidence of an alternative view of the nature of particle acceleration. The frequency of radio emission depends upon the local electron density which decreases with distance from the Sun. Radio observations distinguished fast-frequency-drift type III bursts, produced by a burst of 10-100 keV electrons streaming out from the Sun, from the slow frequency drift of type II bursts, traveling outward at speeds that corresponded to slower, ~1000 km s$^{-1}$, shock waves. Wild, Smerd, and Weiss (1963) suggested that:

(1) electrons were accelerated to produce the type III bursts and that

(2) protons were accelerated at shock waves seen as type II bursts.

Even though *none* of the radio emissions are produced directly by protons, as these authors readily admitted, they had identified two acceleration mechanisms, one dominated by electrons and a separate one with a relative enhancement of protons. The radio observations of two kinds of sources would later be extended to the X-ray realm in the *Skylab* era by Pallavicini, Serio, and Vaiana (1977) who distinguished impulsive and "long-enduring" soft X-ray events. These long-duration X-ray events had been previously associated with coronal mass ejections (CMEs; Sheeley *et al.* 1975). However, we are getting ahead of our story.

A serious problem in determining the site of proton and ion acceleration is that, unlike electrons, they show no characteristic X-ray or radio emission like the bremsstrahlung or synchrotron emission of electrons. Except for the rare γ-ray-line events, which we will discuss, ions undergo "stealth" acceleration and transport that can only be studied by inference or by direct ion measurements in space. Ion acceleration by a shock wave near the Sun, for example, is practically invisible in photons, although the shock itself is not.

We now follow the history of the observations and theories that filled out and extended our understanding of each of these two underlying physical mechanisms.

## 1.1. TYPE III BURSTS AND IMPULSIVE SEP EVENTS

Electrons of ~40 keV of solar origin that can generate type III radio bursts were first observed in space by Van Allen and Krimigis (1965). Subsequently, Lin (1970a, b, 1974) studied events with >22 keV electrons. He found "pure" electron events (*i.e.* events with not-yet-measurable intensities of protons or other ions) and small electron events that were associated with small flares, 20 keV X-ray emission, and type III bursts. Meanwhile, rarer large events with protons (Bryant *et al.* 1962) and relativistic electrons (Cline and McDonald 1968) were found to be distinguished by type II and IV radio emission. Subsequently, Ramaty *et al.* (1980) found two distinct populations of SEP events based upon their e/p ratios, an observation that was confirmed by Cliver and Ling (2007) much later.

However, to follow the evolution of the type-III track and especially the "pure" electron events, we must now consider the discovery of $^3$He-rich events (Hseih and Simpson 1970; Dietrich 1973; Garrard, Stone, and Vogt 1973; Anglin 1975; Serlemitsos and Balasubrahmanyan 1975). It was found that many of these $^3$He-rich events had abundances of $^3$He/$^4$He >0.1, while the corresponding abundance ratio in the corona or solar wind is ~5x10$^{-4}$ (*e.g.* Coplan *et al.* 1984). With improving instruments, a few events were found that even had $^3$He/$^4$He >10 (*e.g.* Reames, von Rosenvinge, and Lin 1985), *more $^3$He than H!* It was also found that these events had increasing enhancements of elements as a function of atomic number for elements up through Fe with Fe/O enhanced by a factor of ~10 relative to abundances in the corona or in large SEP events (*e.g.* Hurford *et al.* 1975; Mason *et al.* 1986, Reames 1988, Reames, Meyer, and von Rosenvinge 1994).

In 1985 $^3$He-rich events were first found to accompany the 10-100 keV electrons that produce type III bursts (Reames, von Rosenvinge, and Lin 1985) and were directly associated with kilometric type III bursts (Reames and Stone 1986). The precise timing of the electron and type III observations led to identification of the impulsive flares associated with these events and motivated a study of X-ray and other properties of the sources (*e.g.* Kahler *et al.* 1987; Reames *et al.* 1988). What was unusual about the source flares that produced these unusual $^3$He-rich events? The only thing that could be found, other than the fact that these were impulsive flares at solar longitudes that were magnetically well-connected to the observer, was that the flares with *smaller* X-ray emission tended to be *more* $^3$He-rich (Reames *et al.* 1988). Otherwise we could find nothing unusual about these flares; they were just typical small C and M class impulsive flares.

An early suggestion for the physics of $^3$He-rich events was advanced by the Fisk (1978) model which proposed selective heating of $^3$He by the resonant absorption of electrostatic ion cyclotron waves produced between the gyrofrequencies of the two dominant species in the plasma, H and $^4$He. Because of the low abundance of $^3$He, each ion could absorb a large amount of energy without significantly damping these waves. However, a second, unspecified mechanism of acceleration was required by the Fisk model. In a later study by Temerin and Roth (1992; Roth and Temerin 1997), streaming electrons were found to produce electromagnetic ion cyclotron (EMIC) waves that could resonantly accelerate $^3$He ions mirroring in the magnetic field, an analogy with acceleration of "ion conics" seen in the Earth's aurora. Heavier elements could be enhanced by second-harmonic absorption of these waves. In a review, Miller (1998) discussed a model that was able to accelerate electrons and heavier ions with the proper abundances by resonating with cascading magnetohydrodynamic waves produced by magnetic reconnection in the flaring region. The cascading waves first resonate with the gyrofrequencies of heavy elements such as Fe, then with Si, Mg, and Ne, then with O, N, C, with He, and finally with H, producing a declining pattern of enhancements. Separately, the streaming electrons produce EMIC waves that accelerate the $^3$He by the Temerin-Roth mechanism. These and most subsequent models *relate impulsive SEP events directly to the fundamental physics of magnetic reconnection and, hence, to flares* (*e.g.* Drake *et al.* 2009). More recently, Liu, Petrosian, and Mason (2004, 2006) have been able to calculate the complex energy dependence of the $^3$He and $^4$He spectra and Drake *et al*. (2009) have shown that magnetic reconnection in a flare leads directly to power-law enhancements as a function of the mass-to-charge ratio, *A/Q*, for ions now measured to the heaviest elements, as we shall see in Section 2.

This evolving evidence led to the idea that the SEPs accelerated in *all* flares were likely to be $^3$He-rich and Fe-rich. In confirmation of this, Mandzhavidze, Ramaty, and Kozlovsky (1999) studied 20 γ-ray-line events and compared relative intensities of de-excitation lines from the products of reactions such as $^{16}$O($^3$He, p)$^{18}$F$^*$ with those from similar reactions involving $^4$He. They found that for the accelerated ions in all of the events, $^3$He/$^4$He could be >0.1 and in some cases it was ~1. *The accelerated ions in the loops of these large eruptive flares were $^3$He-rich even while the SEP ions seen in space (presumably shock-accelerated) were not*. Earlier studies of γ-ray-line events (Murphy *et al.* 1991) had shown that Ne, Mg, Si, and Fe appeared to be enhanced relative to C and O in measurements of the Doppler-broadened γ-ray lines produced by de-excitation of heavy ions of the accelerated "beam" that had collided with H that was at rest in the plasma.

Thus, in large eruptive flares, magnetic reconnection beneath the CME occurs among closed field lines where the reconnected field lines of the loops must also be closed, so that any $^3$He-rich SEPs are confined, eventually scattering into the loss cone and plunging into the low corona where some interact to produce γ rays. One could argue that these events produce large flares *because* the SEPs cannot escape in the time scale of the flare, so that all of the particle energy goes into flare heating. Conversely, impulsive SEP events seen at 1 AU are commonly associated with jets (Wang, Pick, and Mason 2006; Nitta *et al.* 2006), where reconnection between open and closed field lines takes place (*e.g.* Shibata, *et al.* 1992; Heyvaerts, Priest, Rust 1977; Shimojo and Shibata 2000; Aschwanden (2002); Reames 2002), accelerating electrons and ions that have direct access to open field lines. The liberated electrons (and $^3$He) stream outward to produce type III radio bursts. Reconnection sites for jets and the related type III bursts are numerous, but the SEP ion intensities produced in each source are low. A few larger impulsive

SEP events have been associated with narrow CMEs that may represent material ejected upward from the reconnection site of a jet (Kahler, Reames, and Sheeley 2001).

Recently, energetic neutral H atoms have been observed from the solar flare of 2006 December 5 (Mewaldt *et al.* 2009). The neutral atoms are produced by charge exchange of 1.6-15 MeV protons with atoms in the high corona. Timing of the neutral H atoms is related to the flare and these particles arrive about an hour before of the proton onset.

Following Wild, Smerd, and Weiss (1963), we have used type III radio bursts as a signature of impulsive SEP events. The radio emission that is actually observed as a type III burst is generated during electron *transport* out from the source (Kundu 1965), not by the physical mechanism of acceleration. Fast electrons stream out ahead of slower ones to produce a "bump-on-tail" distribution function that is observed to produce Langmuir waves and subsequent radio emission (see Thejappa *et al.* 2012 and references therein). Copious acceleration of electrons in impulsive flares and jets produce thousands of type III bursts annually, but shock waves also accelerate electrons observed in space (see review by Kahler 2007).

Type II bursts are believed (see *e.g.* Ganse *et al.* 2012) to be produced by electrons accelerated at a limited region of the shock surface where the magnetic field lines lie near the plane of the shock, perhaps even intersecting the shock surface in multiple locations. Electrons are accelerated as they drift in the $V_{shock} \times B$ electric field at the shock. Counter-streaming electrons generate counter-streaming Langmuir waves that interact to produce the electromagnetic radiation observed as a type II burst. Since the field lines containing the electrons are soon swept downstream of the shock, the electrons are not likely to escape. If they were able to find a path outward and escape, they might produce a type III burst. Ion acceleration may occur here and elsewhere along the shock. All shocks do not produce type II emission.

## 1.2. TYPE II BURSTS AND GRADUAL SEP EVENTS

From the earliest sounding-rocket observations of abundances of the elements C, N, O (Fichtel and Guss 1961) and Fe (Bertsch, Fichtel, and Reames 1969) in large SEP events, it was clear that these abundances were in some sense "normal" solar abundances, although solar abundances were not yet well known. At first, SEP abundances were thought to be photospheric, but as observation of both SEP and photospheric abundances improved, Meyer (1985) found that SEP abundances were a measure of the corona in which ion-neutral fractionation has caused elements with a high first ionization potential (FIP) to be suppressed by a factor of ~4 relative to those with low FIP. This fractionation occurs during the transport of low-FIP ions and high-FIP neutral atoms from the photosphere to the corona where all elements then become highly ionized and are available to be accelerated democratically by a shock wave. Thus abundances in large SEP events were coronal; the abundances in the impulsive events were enhanced *relative* to these.

CMEs were first observed by Tousey (1973), then on *Skylab* (Gosling *et al*. 1974). Sheeley *et al.* (1975) associated CMEs with long duration X-ray events (LDEs). Kahler *et al*. (1978) were the first to show that all SEP events during the *Skylab* period were associated with a CME or an LDE. A stronger link to SEP events was established with the *Solwind* coronagraph data when *Kahler et al. (1984) found a 96% correlation between large SEP events and fast, wide CMEs*. The important evidence that CMEs drive the shock waves that produce coronal (metric) type II bursts was presented by Cliver, Webb, and Howard (1999), and is now clearly established (Cliver *et al.* 2005; Liu *et al.* 2009; Bain *et al.* 2012).

In a classic study of the onset timing of GLEs, Cliver *et al*. (1982) observed that the *onsets of the relativistic SEPs in these events occurred well after the flash phase of the associated flare*, sometimes by >20 min. They suggested that the acceleration of outbound SEPs began when the shock wave reached open magnetic field lines. Subsequently, Kahler (1994) found that in some large GLEs, the protons up to 21 GeV were accelerated after the shock was above ~5 solar radii.

Lee (1983) applied steady-state diffusive shock acceleration theory to interplanetary shock waves. In this theory, protons streaming away from the shocks generate or amplify Alfvén waves that scatter subsequent particles back and forth across the shock where they gain energy on each transit. However, shock waves in the corona expand through a changing environment that may be poorly approximated by a temporal equilibrium. Ellison and Ramaty (1985) suggested that these factors could be approximated by an exponential rollover of the equilibrium power-law energy spectrum at high energies. Subsequently it was observed that the rollover energy of these spectral "knees" depend upon $Q/A$ of the ion (Tylka *et al*. 2000). Lee (2005) then made a complete analytical study including spectral rollovers and their dependences. The proton-amplified waves have been observed by Viñas *et al.* (1984), Kennel *et al.* (1986), Tsurutani and Gonzalez (1987), Tan *et al*. (1989), and recently by Desai *et al.* (2012).

The onsets of the large "gradual" SEP events are certainly *not* gradual but the events *are* long duration, *i.e.* days, in comparison with hours for impulsive events, partly because of continuing acceleration by the shock. Another major difference is the ionization state of SEP Fe that was found (Luhn *et al*. 1987) to be 14.1 ± 0.2 in gradual events and 20.5 ± 1.2 in impulsive events, the latter elevated by heating or by stripping in flares (DiFabio *et al.* 2008, see also Leske *et al.* 1995; Mason *et al.* 1995; Popecki *et al.* 2002). Even at 200-600 MeV amu$^{-1}$, Fe in large gradual SEP events had an ionization state of 14.1 ± 1.4 (Tylka *et al*. 1995).

The solar longitude distribution of large SEP events was studied by Cane, Reames, and von Rosenvinge (1988). Their interpretation of the time profiles of protons in 285 large SEP events from different solar source longitudes could be understood in terms of the strongest acceleration occurring near the "nose" of a shock that moved radially outward from the Sun with time (Figure 1.1). For observers at solar longitudes to the east of the source, intensities peaked early when they were magnetically well connected to the nose of the shock near the Sun and declined as their connection point swung around the eastern flank of the shock as the shock nose moved radially outward. Observers of a source near central meridian often saw peak intensity when the nose of the shock itself moved over them. Observers to the west of the source saw slowly rising intensities that peaked *after* they passed through the shock and encountered field lines that connected them to the nose of the shock *from behind*. Note that the term "nose" of the shock has been introduced in contrast with the "flanks" and it may poorly represent the complexity, breadth, and extent of the shock surface.

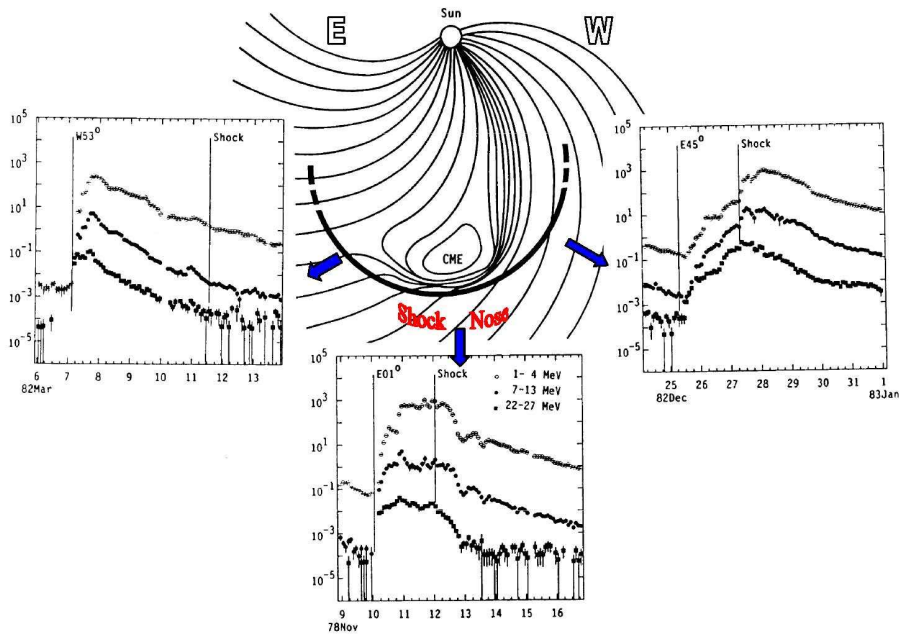

*Figure 1.1.* Typical intensity *vs.* time plots seen for a gradual SEP event viewed from three different solar longitudes relative to the CME and shock wave (see text; after Reames 1999; Cane Reames, and von Rosenvinge 1988).

In an article entitled "The Solar Flare Myth," Gosling (1993) argued against any *causal* relationship between large SEP events and flares, an idea that still remains in the popular perception, and stressed the importance of CMEs and of shock acceleration of SEPs (see also Kahler 1992). He noted that there are magnetically well-connected flares with no associated SEP events and large gradual SEP events with no flares such as the disappearing-filament event of 1981 December 5 (Kahler *et al.* 1986, Cane, Kahler, and Sheeley 1986). He pointed out the strong association of gradual SEP events with fast CMEs and shock waves (*e.g.* Kahler *et al.* 1984, see also Kahler 2001) and of $^3$He-rich SEP events with impulsive flares. The physics of ion acceleration can produce vastly different SEP abundances in flares and shocks, as we shall see; these differences are not visible from electron observations (*e.g.* radio and X-rays). The controversy raised by the Gosling (1993) paper led to an invited discussion from three alternative viewpoints in *Eos* where Hudson (1995) argued that the term "flare" should include the CME, shock, and any related physics, Miller (1995) argued that flares, being more numerous, were a better subject for acceleration studies, and Reames (1995b) argued for the separate study of both flare and shock acceleration of SEPs.

Kahler (2001) showed that peak SEP intensities at different energies were correlated with the CME speeds, as shown in Figure 1.2. While the correlation coefficient was 0.7 for 20-MeV protons, a significant spread of intensities was still evident, showing the presence of other contributing factors, many discussed by Kahler. Subsequently, this study became a basis for comparison of a more recent multi-spacecraft study (Rouillard *et al*. 2012) where individual shock speeds could be measured at the points where the shock intercepted the field lines to each spacecraft (see Section 3.2). Nevertheless, Kahler's study reinforced the growing realization that only the fastest 1-2% of CMEs are able to accelerate SEPs.

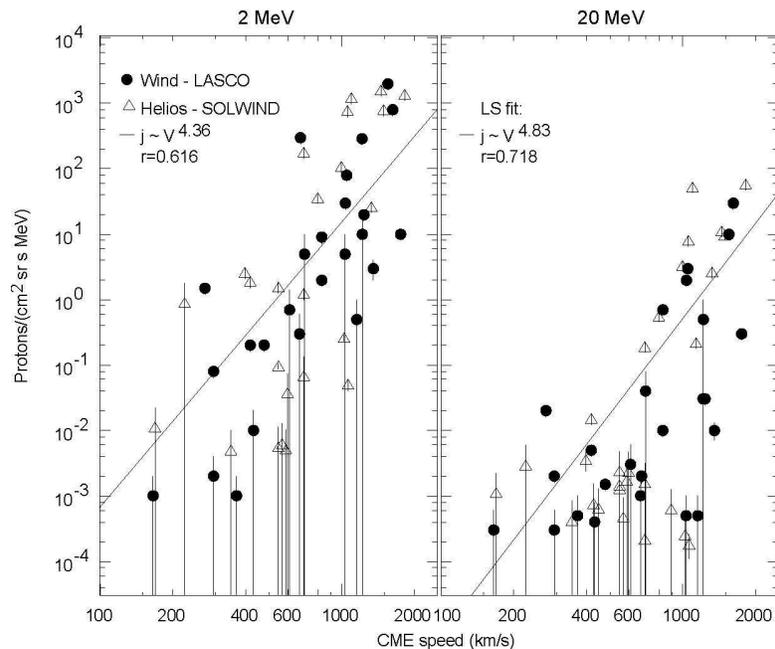

*Figure 1.2* Plots are shown of peak intensity of protons in SEP events at two energies as a function of CME speed (Kahler 2001). Symbols distinguish two combinations of SEP instruments and coronagraphs. Linear least-squares fit lines and correlation coefficients are shown for each proton energy, although the true behavior may not be linear.

Nowadays, GLEs primarily form a convenient class of high-energy SEP events with ~GeV protons. We now measure these GLEs much more completely in space than at ground level, including ~80 MeV amu$^{-1}$ to ~3 GeV amu$^{-1}$ H and He spectra (Adriani *et al.* 2011), onsets (Reames 2009a, b; Reames and Lal 2010), energy spectral shapes and abundances (Mewaldt *et al.* 2012), electrons (Kahler 2007; Kahler *et al.* 2011; Tan *et al.* 2012b) and general properties (Gopalswamy *et* *al.* 2012), often on multiple, spatially separated spacecraft (Berdichevsky *et al.* 2009; Reames and Lal 2010; Reames, Ng, and Tylka 2012).

It is generally believed that most energetic photons produced by SEPs at the Sun come from flares (*e.g.* Hudson and Ryan 1995; Ramaty *et al.* 1995; Vestrand *et al.* 1999; Benz 2008) and that shock-accelerated particles are invisible since they would be mirrored in the converging magnetic field. To interact in the dense corona particle intensities would have to remain high long enough to allow sufficient scattering into the small loss cone. In support of this, Cliver *et al.* (1989) found a poor correlation between intensities of γ-rays and protons in space in most events. However, Vestrand and Forrest (1993) observed γ-ray production extending over ~30$^o$ of the Sun's surface in the large GLE of 1989 September 29 and Ryan (2000) discussed long-duration γ-ray events lasting an hour or more while the flare-associated X-rays died away rapidly. In our modern understanding, it is possible that a reservoir formed behind the CME and shock (see sections 3.1 and 5.4 below) can trap particles that bathe the surface in high-energy particles over a large area for an extended period allowing us to observe shock-accelerated particles from the sunward side, downstream of the shock – an important alternative perspective.

Unlike the acceleration on closed magnetic loops in large eruptive flares, acceleration at shock waves occurs on open magnetic field lines that extend outward into the heliosphere. *Nearly all of the shock-accelerated particles will eventually propagate out beyond the orbit of Earth*. Thus, gradual SEP events are efficient, large, intense, and spatially and temporally extensive. In contrast, impulsive SEP events seen in space are small, weak and compact, but also numerous, where magnetic reconnection includes open magnetic field lines. A number of review articles have contrasted impulsive and gradual SEP events (Ramaty *et al.* 1980, Meyer 1985, Reames 1988, 1990b, 1999; Kahler 1992, 2007; Gosling 1993; Cliver 1996, 2009a, b; Lee 1997, 2005; Tylka 2001; Slocum *et al.* 2003; Reames and Ng 2004; Cliver and Ling 2007; Kahler *et al.* 2011).

## 2. Abundances

Since the work of Meyer (1985), reference abundances in SEP events have been obtained by averaging over large gradual events in the region of ~5-10 MeV amu$^{-1}$ (*e.g.* Reames 1995a). In the lower panel of Figure 2.1 we show the SEP abundances (Reames 1995a, 1999) relative to photospheric abundances from Caffau *et al*. (2011) supplemented by meteoric abundances from Lodders *et al*. (2009), all normalized at H and plotted as a function of FIP. The SEP event-to-event variations in near-neighbor elements such as Mg/Ne are ~20%, so, for a sample of ~43 events, the error in the mean is ~3%, although errors in Fe/O, or He/O, for example, are larger. It is more difficult to determine the abundance of H in SEPs because of some energy dependence in H/He. These and other variations have been discussed in detail by Reames (1995a, 1998, 1999) and energy dependences have been studied by Cohan *et al.* (2007). The variations become larger below ~1 MeV amu$^{-1}$ (Desai *et al.* 2006). Variations are also larger at higher energies, but here isotope resolution up through Fe is also possible (Leske *et al.* 2007). Coronal abundances are also measured with X-ray (*e.g.* Fludra and Schmelz 1999; Landi, Feldman, and Dochek 2007) and γ-ray (Ramaty and Murphy 1987; Ramaty *et al.* 1995) lines and in the solar wind (von Steiger *et al.* 2000, Gloeckler and Geiss 2007). What have changed in recent years are values of element abundances in the photosphere, partly because of improved modeling of the solar atmosphere. Schmelz *et al.* (2012) have recently determined mean abundances at high and low FIP from averages over SEP and solar-wind abundance measurements.

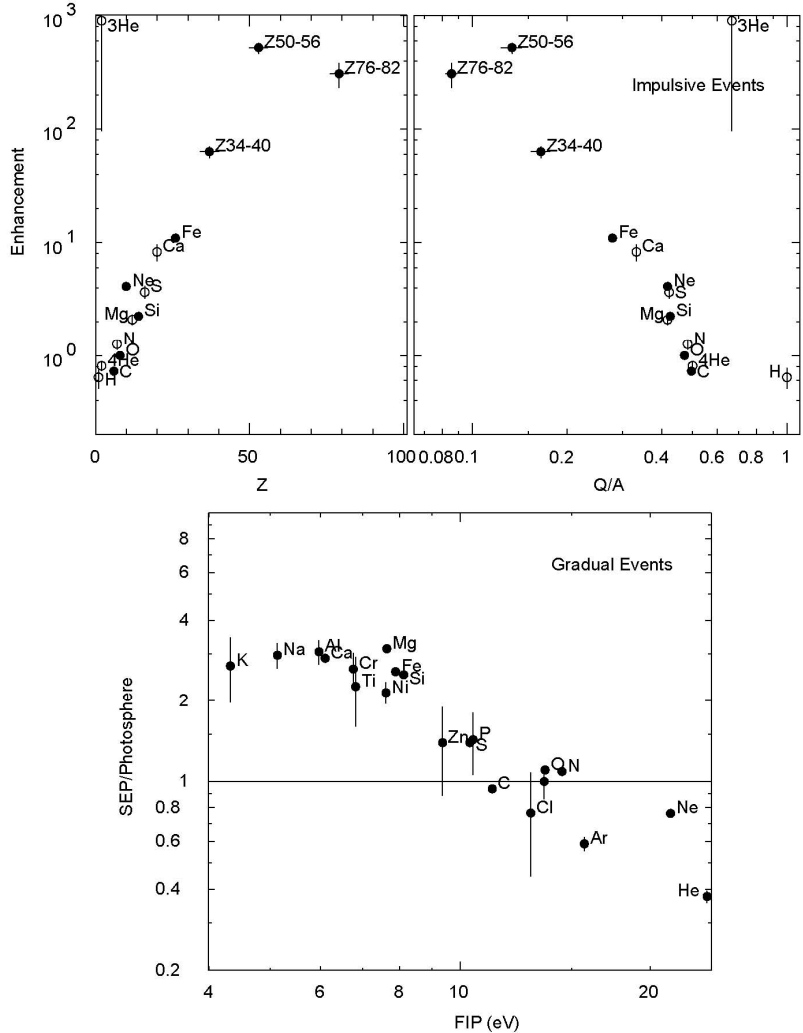

*Figure 2.1.* The lower panel shows average abundance in gradual SEP events relative to photospheric abundances, normalized to H, *vs.* the first ionization potential (FIP) of the element (see text). The upper panels show the average enhancement in impulsive SEP events relative to gradual SEP events *vs. Z* and *Q/A*.

One of the early features that distinguished impulsive and gradual events was differences in abundances of elements up through Fe in addition to isotopes of He. In recent years it has become possible to extend abundance measurements up to the vicinity of the element Pb at $Z=82$ (Reames 2000; Mason *et al.* 2004; Reames and Ng 2004), although not with individual-element resolution above $Z\sim30$. For the impulsive events, the enhancements rise with $Z$ to factors of ~1000, as shown in the upper panels of Figure 2.1 from Reames and Ng (2004). Small variations from a smooth increase are not completely diminished by plotting the increase as a function of the charge-to-mass ratio *Q/A* and no single variable can simultaneously explain the high-Z and $^3$He enhancements, suggesting that separate, but related, mechanisms or wave modes are involved. The magnetic reconnection theory of Drake *et al.* (2009; Knizhnik, Swizdak, and Drake 2011; Drake and Swizdak 2012) does predict a power law enhancement in *A/Q* above a minimum value.

The clean separation of impulsive and gradual events on the basis of abundances, as implied in Figure 2.1, did not last long before Mason, Mazur and Dwyer (1999) found enhancements of $^3$He in some of the large gradual events that were supposed to have only coronal abundances. These values of $^3$He/$^4$He= 1.9±0.2×10$^{-3}$ were small, but were still ~5 times the abundance in the solar wind. The authors suggested that residual suprathermal ions left over from earlier impulsive flares were contributing to the seed population accelerated by the shock. Later, Tylka *et al.* (2001) found that they could explain the energy dependence of Fe/C as well as that of the ionization state of Fe by adding a small fraction of remnant impulsive suprathermal ions to the seed population.

A new and stiffer challenge to the seed-population argument was presented by the abundance comparison of the energy dependence of Fe/C in two otherwise similar events by Tylka *et al.* (2005), as shown in the left panel of Figure 2.2. These two events had occurred at similar solar longitudes and had similar CME speeds. The presumed behavior of the spectra of the seed population is shown in the right panel of the Figure 2.2. Since the impulsive suprathermal spectrum is likely to be harder, it will dominate at higher energies so the Fe/O abundance ratio in the seed population increases with energy, as shown in the inset. A significant factor affecting the abundances will occur for quasi-perpendicular shock waves where the injection threshold may be higher since ions must be fast enough to catch up from behind the shock along the field ***B*** despite a large angle $\theta_{Bn}$ between ***B*** and the shock normal (Tylka *et al.* 2005, 2006).

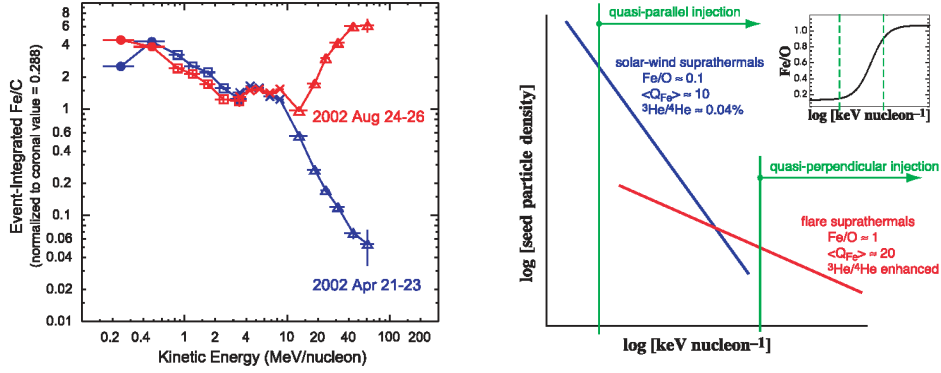

*Figure 2.2* The left panel compares the energy dependence of Fe/C for two gradual events that are otherwise similar in their properties (Tylka *et al.* 2005). The right panel shows hypothetical spectra of suprathermal ions where different injection thresholds will yield different abundance ratios. Clearly, one could not use Fe/C to distinguish impulsive and gradual SEP events above ~10 MeV amu$^{-1}$.

The comprehensive theory of Tylka and Lee (2006) helps to explain abundance variations caused by variations in the selection of the seed population and their affect on the accelerated spectra. Figure 2.3 shows typical variations in Fe/O as a function of energy for different values of *R*, the ratio of O in the impulsive suprathermal component to that in the coronal component of the seed population. The accelerated energy spectra are assumed to be modified at high energy by the Ellison-Ramaty (1985) exponential factor, $\exp(-E/E_{0i})$, to account for a finite acceleration time, where $E_{0i}$ is proportional to $Q_i/A_i$ for the *i*-th species, so that the intensity $j_i(E)=kE^{-\gamma}\exp(-E/E_{0i})$. $E_{0i}$ also depends upon $\sec\theta_{Bn}$ to account for the shorter acceleration time at quasi-perpendicular shock waves (Lee 2005; Tylka and Lee 2006). Thus,

$$E_{0i} = E_0\left(Q_i/A_i\right)\times\left(\sec\theta_{Bn}\right)^{2/(2\gamma-1)} \qquad (2.1)$$

where $E_0$ is the proton knee energy for a quasi-parallel shock wave, and *γ* is the energy power-law exponent. The factor $Q_i/A_i$ is different for Fe and for O in the impulsive and coronal seed particles. This makes the energy of the spectral break depend strongly upon the composition of the seed population and on $\theta_{Bn}$ so the ratio of SEP Fe/O can break upward or downward depending upon *R* (Figure 2.3).

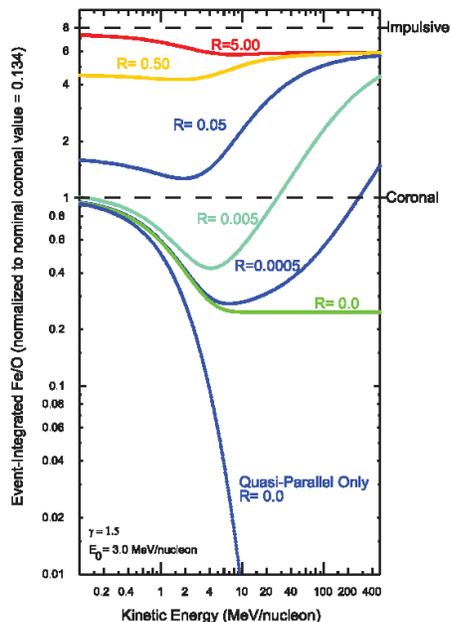

*Figure 2.3.* Typical theoretical variation of the abundance ratio Fe/O, relative to the coronal value, as a function of energy, is shown for different values of the impulsive suprathermal fraction, *R*, in the seed population (Tylka and Lee 2006). Small values of *R* can lead to large changes in Fe/O at energies above the spectral knee.

At energies above ~5 MeV amu$^{-1}$, when Fe is enhanced, relative to O, elements with intermediate values of *Q/A*, *i.e.* Ne, Mg, Si, *etc.*, have intermediate enhancements, and similarly for suppressions. Thus the theory of Tylka and Lee (2006) explains the power-law dependence of abundances on *Q/A* observed much earlier by Breneman and Stone (1985), as well as the equivalent "mass-biased" SEP event abundances of Meyer (1985) that led to his averaged or "mass-unbiased" SEP estimate of solar coronal abundances.

In a short time we have moved from an apparent blurring of the simple abundance patterns of impulsive and gradual events to a theoretical model of the seed population that explains the abundance variations and leaves us with the same two primary physical acceleration mechanisms that we had at the beginning.

Another feature of large SEP events relates to the cause of abundance variations with time early in the events. Early increases in Fe/O (but not in $^3$He), for example, were erroneously ascribed to the presence of material accelerated in the impulsive flare preceding the shock acceleration, especially in magnetically well-connected events (*e.g.* Reames 1990a, Cane *et al.* 2003). In Section 5.1 we will see that the very first ions that arrive from an event are minimally scattered and have coincident onsets. However, the initial abundance ratio, and the subsequent increase in intensity depends upon interplanetary scattering and, hence, upon *Q/A*, so that Fe scatters less and increases earlier than O, in most cases, causing the early maximum in Fe/O. In general, the initial behavior of ratios like

Fe/O or $He^4$/H can depend upon wave amplification by streaming protons early in large SEP events (Reames, Ng, and Tylka 2000) producing rising or falling $He^4$/H ratios (see also Ng, Reames, and Tylka 2003). These differences in transport can also produce variations in Fe/O with pitch angle so that Fe/O can be higher for particles flowing sunward than for those flowing outward along the magnetic field (Reames and Ng, 2002). Recent observations from *Wind* and *Ulysses* also have shown that early increases in Fe/O are unrelated to the solar source longitude and result entirely from differential transport (Tylka *et al*. 2012).

Only a brief glance at the left panel of Figure 2.2 or at Figure 2.3 is required to make clear that there are *large* abundance variations at high energies that are evidently caused by the presence of impulsive suprathermal ions, energy spectral knees, and differences in $\theta_{Bn}$. *Clearly, one cannot easily study coronal abundances or even distinguish between impulsive and gradual events using Fe/O at energies above ~10 MeV amu$^{-1}$.* Below ~1 MeV amu$^{-1}$, ions take many hours to propagate from the Sun, and a large fraction of the observed ions are accelerated locally from the solar wind or other ambient plasma. This means that particle transport effects are large and that coronal abundances are unlikely to dominate. Thus, it has been suggested (A. J. Tylka, private communication) that the region near a few MeV amu$^{-1}$ is the "sweet spot" for abundance studies of a generic nature. Abundances at lower energies are ideal for studies of $^3$He and $^4$He spectra (*e.g.* Mason *et al.* 2002; Liu, Petrosian, and Mason 2006) and of local *in situ* shock acceleration (*e.g.* Desai *et al.* 2003, 2004) and at higher energies for the study of shock spectral evolution and its dependence upon $\theta_{Bn}$ (*e.g.* Tylka and Lee 2006). However, the "sweet spot" in the region of a few MeV amu$^{-1}$ best preserves information about the underlying source abundances.

Finally, we cannot leave the subject of abundances without noting the separation of gradual and impulsive SEP events based upon the e/p ratio by Ramaty *et al.* (1980) and recently by Cliver and Ling (2007). These studies go directly to the heart of supporting the predictions by Wild, Smerd, and Weiss (1963). However, electrons, like ions, have both flare and shock sources (see the extensive review of SEP electrons by Kahler 2007)

# 3. Multi-Spacecraft Observations

## 3.1 RESERVOIRS

As we have described them, gradual SEP events are an extensive spatial phenomenon and the measurement of many events from a single location near Earth is a poor substitute for measuring a single event from multiple locations. An excellent example of this is shown in Figure 3.1 where the 1979 March 1 SEP event was observed by three spacecraft (Reames, Kahler, and Ng 1997; Reames 2010). The paths of the spacecraft through the expanding CME are shown in the lower panel, although, of course, in reality the spacecraft nearly remain fixed as the CME expands past them. *Helios 1* encounters the event near central meridian and sees an intensity peak of 3-6 MeV protons near the time of shock passage and subsequently passes through the helical magnetic cloud (MC; see Burlaga *et al.* 1981). As each of the other two spacecraft pass through the shock, their intensities reach a peak then begin to track the intensity at *Helios 1* after they enter the "reservoir" region. At the time interval labeled R, all three spacecraft have entered the reservoir and have nearly identical spectra as shown in the right-hand panel in the figure. This uniform intensity and spectral shape in the reservoir persists while the overall intensity declines with time as the volume of the "magnetic bottle" expands adiabatically (for theory, see appendix of Reames, Barbier, and Ng 1996).

Reservoirs were first reported from the *Interplanetary Monitoring Platform* (IMP) *4* and *Pioneer 6* and *7* observations of ~20 MeV protons spanning ~180$^o$ in solar longitude by McKibben (1972). Twenty years later reservoirs were seen extending over 2.5 AU radially between IMP 8 and *Ulysses* by Roelof *et al*. (1992). The spectral invariance as a function of time and space from ~1-100 MeV was described by Reames, Barbier, and Ng (1996) and by Reames. Kahler, and Ng (1997). Reservoirs extend to *Ulysses* at heliolatitudes up to >70$^o$, N and S (Lario 2010), and they are also seen in electrons (Daibog, Stolpovskii, and Kahler 2003).

Historically, it was common to represent multi-spacecraft observations of SEP events by plotting the peak intensity at each spacecraft as a function of longitude, often showing a strong apparent gradient with longitude. However, for the SEP event in Figure 3.1, at the time when IMP 8 intensity reaches a peak, there is no longitude variation at all. Furthermore, the intensity maxima at *Helios 2* and IMP 8 are just equal to the reservoir intensity at the time each spacecraft enters it, a *variation only in time*, *not space*. Differences in the *time* of maximum intensity are important for correct interpretation so that reservoirs are not overlooked.

The extent of a reservoir may also be determined from a single spacecraft by normalizing the intensity profiles of particles of different energies at a single point in time. The different energies remain normalized for an extended time in the reservoir or "invariant spectral region" (Reames, Kahler, and Ng 1997). Figure 3.2 shows examples of this for two different SEP events. Note that for the 1995 October 20 event, shown in the left panel of the figure, this reservoir begins ahead of the time of shock passage for this event from a source at W55°. This is why the reservoir region, shaded with horizontal red lines in Figure 3.1, extends beyond the shock on its eastern flank. The point at which the reservoir begins in the left panel of Figure 3.2 corresponds to a tangential discontinuity in the magnetic field, indicating that the field structure is more complex than is indicated in Figure 3.1. Several CMEs and shocks were observed to precede the onset of this event adding to the complexity of the magnetic field and forming magnetic boundaries prior to the arrival of any particles from the October 20 event. However, the outer bound of many well-defined reservoirs at <30 MeV occurs at the magnetic compression region downstream of the CME-driven shock on its central and western flanks.

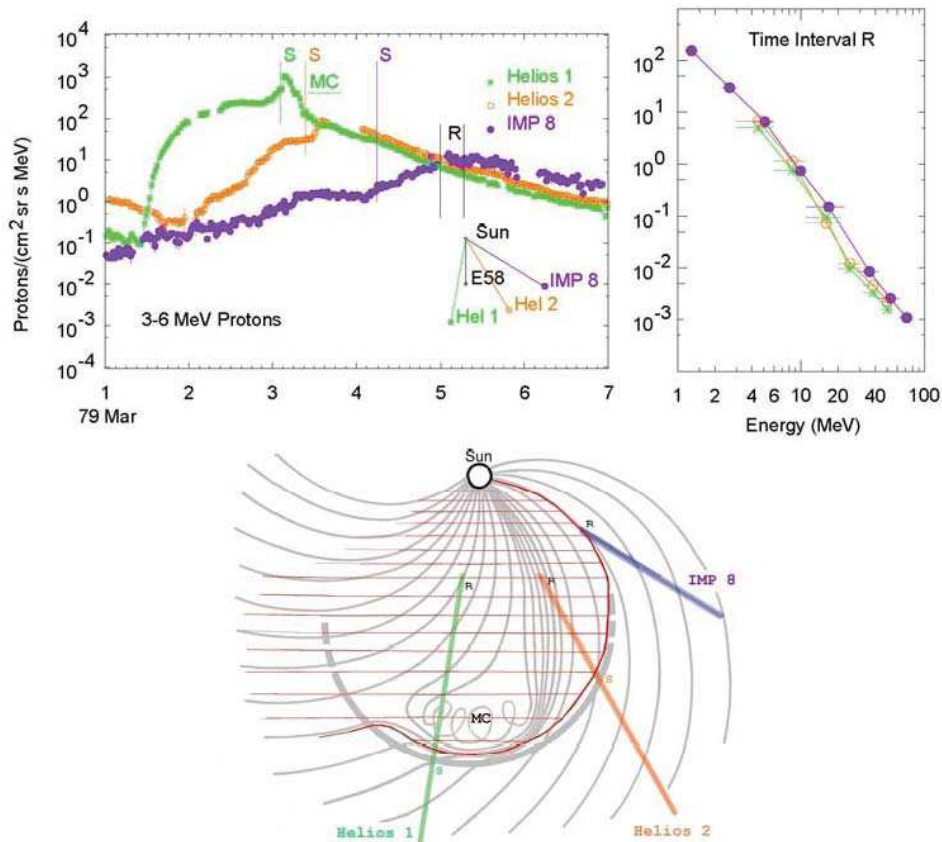

*Figure 3.1* Intensity-time profiles for protons in the 1979 March 1 event at 3 spacecraft are shown in the upper left panel with times of shock passage indicated by S for each. Energy spectra in the "reservoir" at time R are shown in the upper right panel while the paths of the spacecraft through a sketch of the CME are shown below (Reames 2010).

Note that *leakage of particles from reservoirs must be minimal*. For example, if high-energy ions could flow rapidly to the boundary and leak out, the reservoir spectrum would steepen with time, contrary to the commonly observed invariance.

While reservoirs are normally observed to be a phenomenon of gradual SEP events, one could argue that they must form for impulsive events as well. While there is no shock associated with impulsive SEP events, particles may certainly be detained by magnetic structures from preceding CMEs. In recent multi-spacecraft observations of $^3$He-rich events (Wiedenbeck *et al*. 2011), low intensities of ions have been seen at distant longitudes (~136°) beginning almost ~16 hrs after those at magnetically well-connected spacecraft. Clearly these remote ions have traveled a great distance in this time and are no longer on field lines that thread the flare. We will discuss this slow cross-field transport in a reservoir in Section 4.2. Once a magnetic bottle is formed, energetic particles from nearly any sources can fill it, but the process of filling (or emptying) is quite slow.

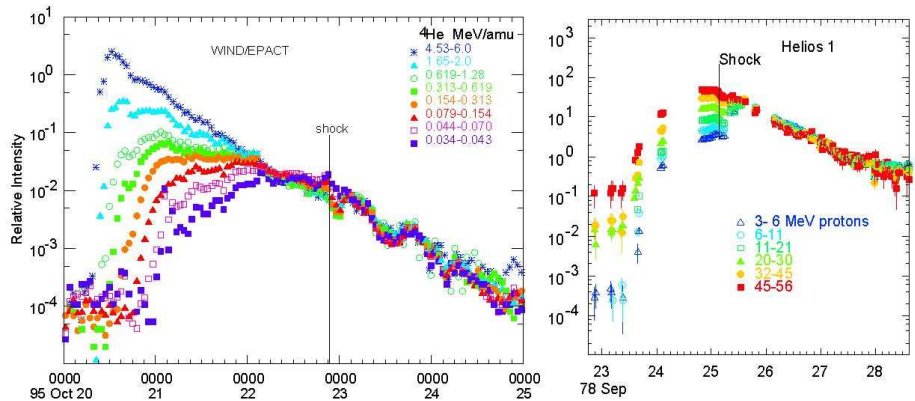

*Figure 3.2.* An invariant spectral region occurs when particle intensities at different energies maintain their normalization as a function of time as shown for different species in two different events (Reames *et al*. 1997a; Reames, Kahler, and Ng 1997).

### 3.2 THE NOSE OF THE SHOCK

Multi-spacecraft observations can also provide information on the solar longitude of the maximum SEP intensity near the time of shock passage, *i.e.* the longitude of the "nose" of the shock as well as the angular breadth of this region of strong acceleration. Panels in Figure 3.3 show intensity-time profiles of protons at three energies and electrons at one for the 1978 January 1 event as observed at 4 locations in space shown in the right-hand panel. During this SEP event, *Helios 2* and IMP 8 see an event near central meridian and protons up to ~30 MeV reach a peak just after the shock passage.

As is often the case, evidence of a reservoir is seen at different times for different energies in this event. For electrons and for protons >100 MeV, equal intensities at *Helios 1*, *Helios 2*, and IMP 8 are seen beginning about midday on January 2. For protons <30 MeV, all four spacecraft have comparable intensities beginning midday on January 4.

Most interesting in the 1978 January 1 SEP event is the intensity peak at energies below about 30 MeV at *Voyager 2* (and also at nearby *Voyager 1*) at the time of local shock passage at 1.95 AU. At low energies the shock peak is just as intense at *Voyager* as it was at *Helios 2*, and IMP 8. On January 6 and 7 intensities of 6-11 MeV protons are the same at all four spacecraft (electrons and higher energy protons have returned to spacecraft-dependent background by this time). The *Helios* era provides a wealth of unique distributions of spacecraft that are unlikely to be duplicated in the near future. Numerous SEP events during this period have provided a test of our new understanding of SEP spatial distributions and, given modest support, will continue to provide new value for years to come.

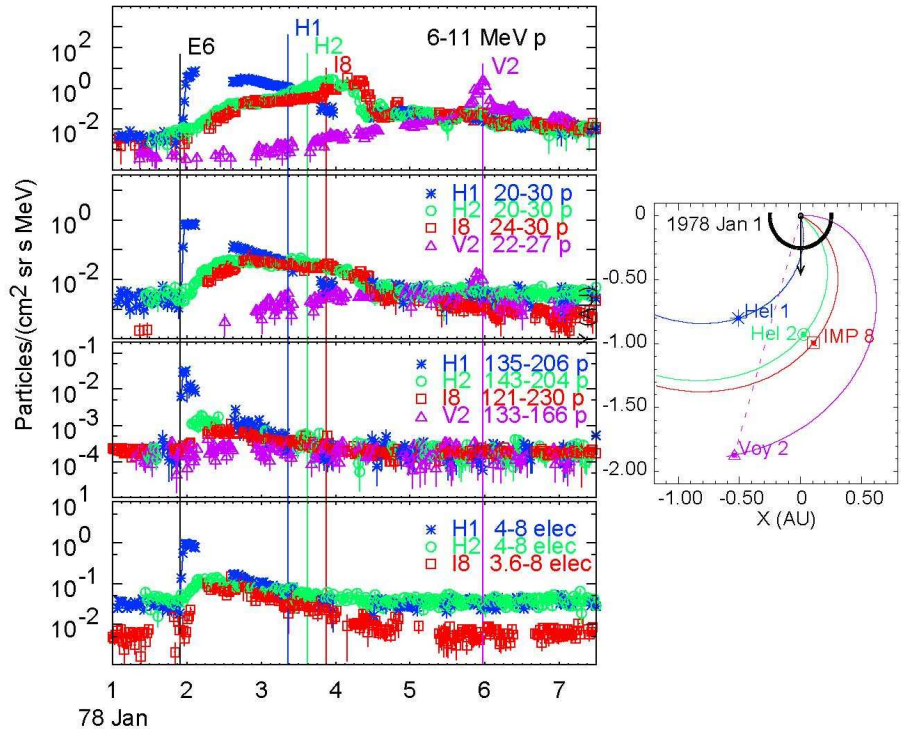

*Figure 3.3.* Intensities of protons and electrons of the energies listed are shown as observed by 4 spacecraft with the spatial distribution shown in the right-hand panel (after Reames, Ng, and Tylka 2012). Vertical lines indicate the time and longitude of the flare (E6) and the times of shock passage at each spacecraft. The spacecraft are *Helios 1* (blue asterisks), *Helios 2* (green circles), IMP 8 (red squares), and *Voyager 2* (violet triangles).

Observations from the *Solar TErrestrial Relations Observatory* (STEREO) add one important new factor to our study of SEP spatial distributions. The coronagraphs on STEREO can *image the shock wave* and allow its spatial structure, speed, and location to be studied as a function of time. A collection of STEREO observations is shown in Figure 3.4. In the upper right panels of the figure, images that include the shock position are shown, while the left panels show a reconstruction of the evolution of the CME and shock (of a different event) in a constant latitude plane at three times, with the magnetic field line to Earth shown as a dashed line. This series shows that the shock intercepts the field line to Earth quite late in the event at a time which corresponds to a late SEP onset.

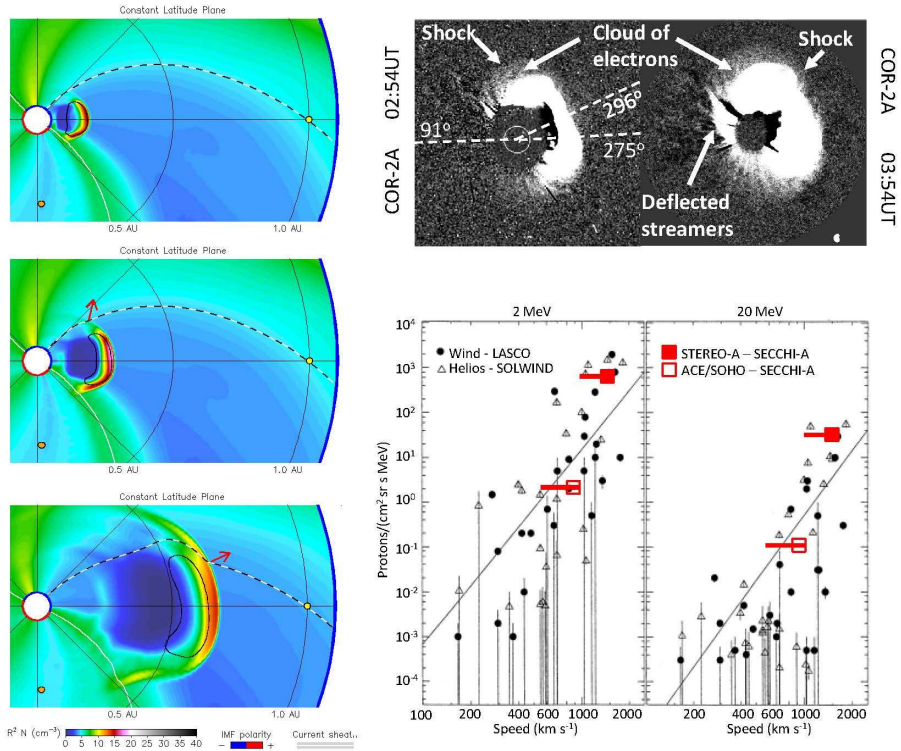

*Figure 3.4.* The left-hand panels show a *STEREO* reconstruction of the time evolution of the CME and shock as it strikes the field line to Earth at ~0.5 AU in the 2010 April 3 SEP event (Rouillard *et al* 2011). The upper right panel shows *STEREO* coronagraph images of features of the CME of 2011 March 21 (Rouillard *et al.* 2012). The lower right panels show the location of the SEP peak intensity *vs.* shock speed measurements of 2 and 20 MeV protons at *STEREO A* and at Earth on the intensity *vs.* CME speed plots from Kahler (2001) shown in Figure 1.2 (Rouillard *et al* 2012).

The STEREO coronagraphs can also determine the shock speed where it strikes the magnetic field line to Earth or to one of the STEREO spacecraft. The lower right panels in Figure 3.4 map the peak proton intensity *vs.* shock speed at two observation points (Rouillard *et al.* 2012) onto a plot of peak proton intensity at 2 and 20 MeV *vs.* CME speed for entire events from Figure 1.2.

Much of the scatter of points in the plot of intensity *vs.* CME speed may occur because both shock speed and the peak intensity vary with position along the shock. Knowing the shock speed and SEP intensities on individual flux tubes connecting each observer to the shock may improve our knowledge of the functional form of this relationship. Figures 3.1 and 3.3 both show typical SEP events where the peak intensity varies with longitude at any energy, and the STEREO measurements show variations in shock speed with longitude. Thus the correlation between SEP intensity and a single CME or shock speed describes only the general behavior of a more complicated situation.

Finally, we note the wide latitude expanse of shock-accelerated near-relativistic electrons. In the 2000 November 8 SEP event these electrons were observed near Earth and on *Ulysses* at 2.4 AU and at a heliographic latitude of ~80° S (Agueda *et al.* 2012).

## 4. Particle Transport

Particle transport played a major role in early attempts to understand SEP events, some of it counterproductive. Why are gradual events gradual? How do particles reach solar latitudes and longitudes so far from a flare? Do particles cross magnetic field lines? In more modern terms, why are reservoirs so uniform?

In the early days, before the discovery of CMEs (in 1973) and the importance of shock waves, SEPs were commonly thought to originate in flares (discovered in 1859) so that all of their subsequent variations were believed to come either from transport or from an arbitrarily assumed injection time profile at the flare site.

### 4.1 PARALLEL TRANSPORT

In early studies of SEP events, impulsive events were *only* observed in the form of "scatter-free electron events" (now $^3$He-rich events). All other events, which we now call gradual, were thought to be gradual because of slow diffusion of the particles in interplanetary space with a fixed, pre-existing parallel scattering mean free path $\lambda_\parallel$ ~0.1 AU. In fact, a review by Palmer (1982) established the "Palmer consensus" that $\lambda_\parallel$ was in the range 0.08 – 0.3 AU. The association of $^3$He-rich events with the scatter-free electron events raised an important question. Did the difference between impulsive and gradual events lie in acceleration, transport, or both?

This question arose most clearly when Mason, Ng, Klecker, and Green (1989) published the figure shown here as Figure 4.1. In the middle of the slow decline of the 0.6-1.0 MeV amu$^{-1}$ H and He intensities in a small gradual event, which began on 1978 October 20, suddenly there arose an impulsive event on October 23$^{rd}$, seen most clearly in He, that was minimally affected by interplanetary scattering. A typical fit of the Fokker-Planck equation (including diffusion, convection and adiabatic deceleration) to the early portion of the proton intensity profile is shown with a radial scattering mean free path $\lambda_r$=0.11 AU. Clearly, the impulsive event on October 23 must have $\lambda_r$>1 AU. How could scattering depend upon the source?

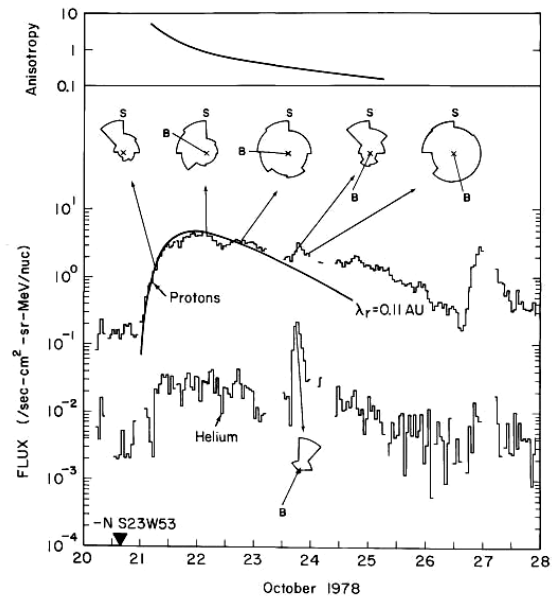

*Figure 4.1.* The small impulsive event on October 23$^{rd}$ appears during the slow decline of the gradual event beginning on the 20$^{th}$. Does diffusion control SEP profiles? (Mason *et al.* 1989)

If we revisit observations at times just prior to the event of 1978 October 20, we find a series of large SEP events on October 8, 9, 13, and 17, with evidence of a final CME passing on October 18 with a strong compression of the plasma density and magnetic field ahead of it. When particles flowed out from the SEP event on the 20$^{th}$, they encountered this last CME, then at ~1.5 AU, and began to fill the magnetic flux tubes that contained them. That is, these particles began filling a reservoir. When we look at the SEP data from ISEE 3 and IMP 8, there appears to be an invariant spectral region running from midday on the 21$^{st}$ through the 26$^{th}$. Thus, *the slow decline in particle intensity arose from the slow expansion of the volume of this reservoir, not from interplanetary scattering*!

When we look back at the data for many of the SEP events that led to the Palmer (1982) consensus, *we often see reservoirs, not slow transport*. Note that the fault is not in the use of the Fokker-Planck or Boltzman equations (*e.g.* Mason *et al.* 1989) to fit the observations of *small* gradual events. Addition of a reflective outer boundary beyond 1 AU (especially one that is moving radially at the solar-wind speed), *in lieu* of a small $\lambda_\parallel$ or $\lambda_r$, can simulate a reservoir. However, wave generation by the streaming protons may also be a factor in *large* SEP events (Ng, Reames, and Tylka 2003). At solar maximum, it is estimated (Reames 1999) that CME remnants have an average radial spacing of ~1 AU in any random direction from the Sun. Reames and Ng (2002) found that a reflective outer boundary at ~1.8 AU was required to explain differences in the angular distributions of Fe and O in several large gradual SEP events, and evidence for preceding CMEs was found. Simulations by Ng, Reames, and Tylka (2003) used an outer boundary, initially at 2.0 AU, which moved out at the solar-wind speed.

Thus, it is not uncommon to have an old CME beyond 1 AU at the onset of a new SEP event at the Sun. In some cases, reflected particles with pitch angle cosine, $\mu$~ -1 can be seen returning sunward a short time after outflowing particles with $\mu$~1 have streamed past (Tan *et al.* 2009, 2012a). Particles are not seen at $\mu$~0 during this time and evidence of a loss cone near $\mu$= -1 may be seen in the reflected particles.

From the foregoing it may seem that interplanetary scattering is actually initially minimal for most SEP events, both impulsive and gradual. However, SEPs streaming outward from large gradual events amplify outward (anti-sunward) Alfvén waves by orders of magnitude (and damp sunward Alfvén waves). This will cause considerable scattering of the SEPs that follow (Ng, Reames, and Tylka 2003) and intensities may even reach the streaming limit (see Section 5.2). Note that particles streaming sunward will damp these outward Alfvén waves. Thus

*SEPs can create (or destroy) their own scattering environment so that $\lambda_\parallel$ varies in both space and time.* At strong shocks, $\lambda_\parallel$ becomes very small, even approaching the Bohm limit where it becomes comparable with the proton gyroradius, $\rho_p$.

Electrons behave differently from ions because they resonate with other wave modes in the interplanetary plasma. They may be scatter free ($\lambda_\parallel \geq 1$ AU) or diffusive (Tan *et al.* 2011). Frequently an energy transition region occurs between non-relativistic electrons, which are usually scatter free, and relativistic electrons, which are diffusive. This complicates the use of electrons to determine onset timing of SEP events using velocity dispersion. This problem caused Kahler, Krucker, and Szabo (2011) and Kahler, Haggerty, and Richardson (2011) to use the time between the type III radio burst onset and the arrival of non-relativistic electrons from impulsive events to determine field-line lengths in magnetic clouds.

## 4.2 CROSS-FIELD TRANSPORT

In the early history of SEP studies, a *fast* transport process called "coronal diffusion" of SEPs from a flare source was often assumed as an alternative to shock acceleration for explaining the wide longitude distribution observed early in gradual SEP events (see *e.g.* Wibberenz 1979). With improving measurements, Mason, Gloeckler, and Hovestadt (1984) found that SEP abundances could not be understood from cross-field transport in the corona but were explained by independent shock acceleration at each longitude. Despite the demise of coronal diffusion from a variety of evidence, the uniformity of intensities in reservoirs leaves us with the question: to what extent and how do particles cross field lines?

In this context we consider two examples. First, Mazur *et al.* (2000) examined the arrival of particles of different energies from impulsive and gradual SEP events as shown in Figure 4.2. Gaps in the arrival of ions from the impulsive

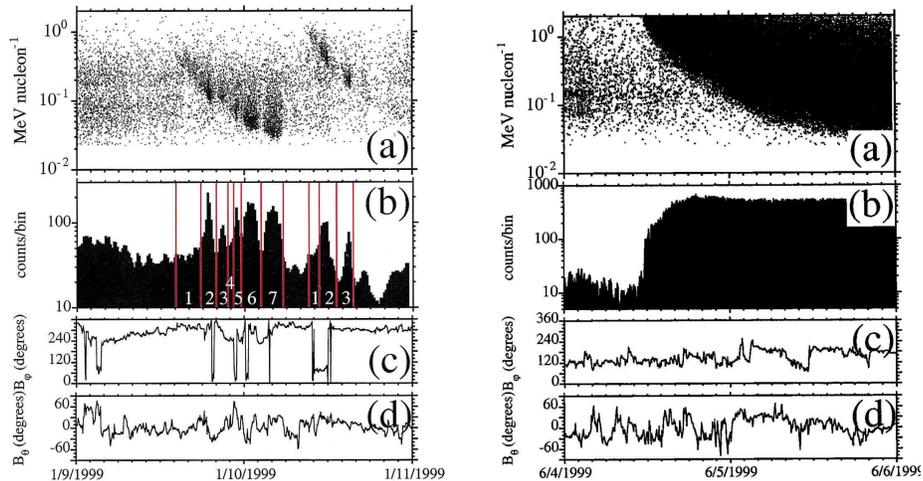

*Figure 4.2.* Panels (a) show the arrival of individual ions *vs.* energy from an impulsive event (left) and from a gradual event (right). Lower panels show (b) the ion count and the (c) azimuthal and (d) polar angles of ***B*** (Mazur *et al.* 2000)

event are produced when the spacecraft samples magnetic flux tubes that do, or do not, intercept the compact impulsive SEP source at the Sun. For the gradual event, all flux tubes are filled at the shock. Subsequent work with impulsive events (Chollet and Giacalone 2011) *has shown the gap boundaries to be extremely sharp* indicating that these ions *do not cross* from filled to neighboring empty flux tubes in their journey out to 1 AU; *i.e.* $\lambda_\perp \gg 1$ AU. The initial longitude spread (*e.g.* half width at half maximum) of field lines that actually thread the source was estimated from *promptly* arriving impulsive SEP events as ~$20^o$ (Reames 1999).

A second, alternate example of cross-field transport is provided by the transport of low-energy anomalous cosmic rays (ACRs) into magnetic clouds (Reames, Kahler, and Tylka 2009; Reames 2010). In magnetic clouds (Burlaga *et al.* 1981) behind CMEs, it is common to observe counter-streaming electrons that are believed to flow up into each end of the field lines from the tail of the thermal electron distribution in the corona (Gosling *et al.* 1987; Shodhan *et al.* 2000). Equal electron flow in either direction is generally taken as evidence that the field lines are closed loops, *i.e.* they are connected directly back to the corona at both ends. There are ~30 examples of CMEs with magnetic clouds that occur during the solar minimum period from 1995-1998 (Shodhan *et al.* 2000). Most of these CME are too slow to drive shock waves that accelerate SEPs above 1 MeV.

When we examine energetic particles associated with these magnetic clouds, as in the example in Figure 4.3, we see element abundances associated with ACRs from the outer heliosphere (He/O ~1, O/C ~20). During the time it took for the cloud to expand to 1 AU, *i.e.* 3-4 days, it has filled with ACRs to an intensity level of ~95% of the ambient level outside the cloud (Reames 2010). How did the ACRs get into the "closed" magnetic clouds?

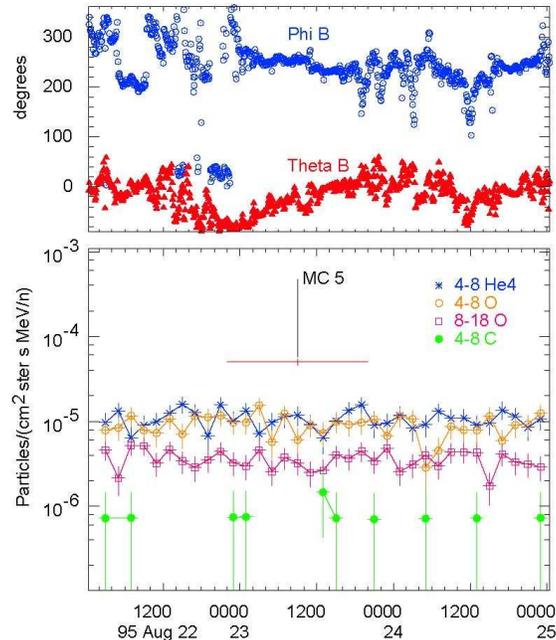

*Figure 4.3* In the upper panel, magnetic field polar and azimuth angles show the rotation of a magnetic cloud (MC 5) which also has bidirectional electron flows (Shodhan *et al.* 2000). The lower panel shows ion intensities with abundances typical of anomalous cosmic rays (He/O ~1, O/C~20) from the outer heliosphere (Reames, Kahler, and Tylka (2009).

It is well known that galactic cosmic rays (GCRs) also fill magnetic clouds, without associated shock waves, to a level only 0.5-2.5% below the ambient level (Zhang and Burlaga 1988). These relativistic particles are affected by gradient and curvature drifts which can produce GCR drift times as short as 2.3 hr (Krittinatham and Ruffolo 2009). However, since drifts scale as the square of velocity, this would be ~10 days for ~4 MeV amu$^{-1}$ $^4$He, so drifts are an unlikely explanation.

To simultaneously explain the presence of both bidirectional electrons and ACRs, we must assume that most of the field lines in magnetic clouds are actually open while some few are closed, with mixing on a fine spatial scale (Reames 2010). In addition to probable reconnection behind the CME there may already be a network produced by random walk of the field lines in the corona (Jokippi and Parker 1969). Comparison of the times for reservoir formation by electrons and protons (Daibog, Stolpovskii, and Kahler 2003; Reames, Ng, and Tylka 2012) shows a dependence upon particle velocity (rather than rigidity) that suggests particles must wander for long distances to propagate significantly in longitude. In fact, the ~4 MeV amu$^{-1}$ $^4$He ACR ions travel ~16 AU day$^{-1}$. Thus we might crudely estimate that $\lambda_\perp \sim 10$ AU (within a factor perhaps as large as ~2) as a compromise between one or two day of ACR travel ($\lambda_\perp < 16$ to 32 AU) and the selectively filled flux tubes ($\lambda_\perp \gg 1$ AU) found by Mazur *et al.* (2000) and by Chollet and Giacalone (2011). This would be adequate to keep CMEs, magnetic clouds, and reservoirs filled with ACRs or with SEPs as they expand radially. However, this process is perpendicular transport, *not perpendicular diffusion*. The particles do not spread laterally at each radius, but they fill each flux tube, finding isolated crossover points at varying radii to fill the next flux tube and eventually explore every nook and cranny of the field network. Like the ACRs, the particles from impulsive SEP events are also substantially delayed (~16 hrs) and attenuated by spreading through a reservoir to distant longitudes (Wiedenbeck *et al.* 2011).

Obviously the cross-field transport affects particle spectra, which tend to become uniform across the reservoir. It also affects abundances in a similar manner. Thus the unsuccessful attempt by Kahler, Tylka, and Reames (2009) to distinguish SEP abundances accelerated in regions with different solar-wind speeds was probably undermined by this slow cross-field transport. We should also note that if cross-field transport were rigidity dependent, we might expect to find regions in MCs, behind large SEP events, or at reservoir edges that were Fe-rich, for example, because Fe has a higher rigidity than O, at a given velocity. This is actually an extension of the argument used by Mason, Gloeckler, and Hovestadt (1985) to argue against coronal diffusion.

Note that the cross-field transport that we find is extremely slow (~ a day) relative to that that would be required for "coronal diffusion" to explain the longitude spread of SEP onsets (minutes). However, in contrast, in the strongest shocks, extreme turbulence may disrupt mean field directions so that $\lambda_\perp \sim \lambda_\parallel \sim \rho_p$.

## 5. Phases of a Large Gradual SEP Event

Multi-spacecraft studies have shown us that a large gradual SEP event, viewed from a longitude near the nose of the shock, has a temporal structure like the event shown in Figure 5.1. Here we can identify 4 distinct phases during which *different physical processes* modify the intensities, abundances, and energy spectra. The phases, noted by Lee (2005) and shown at the bottom of Figure 5.1, are: (1) the onset, (2) the plateau, (3) the shock peak, and (4) the reservoir. In subsequent sections we consider each of these phases in turn. In the first and last cases, the physical processes apply to both impulsive and gradual events.

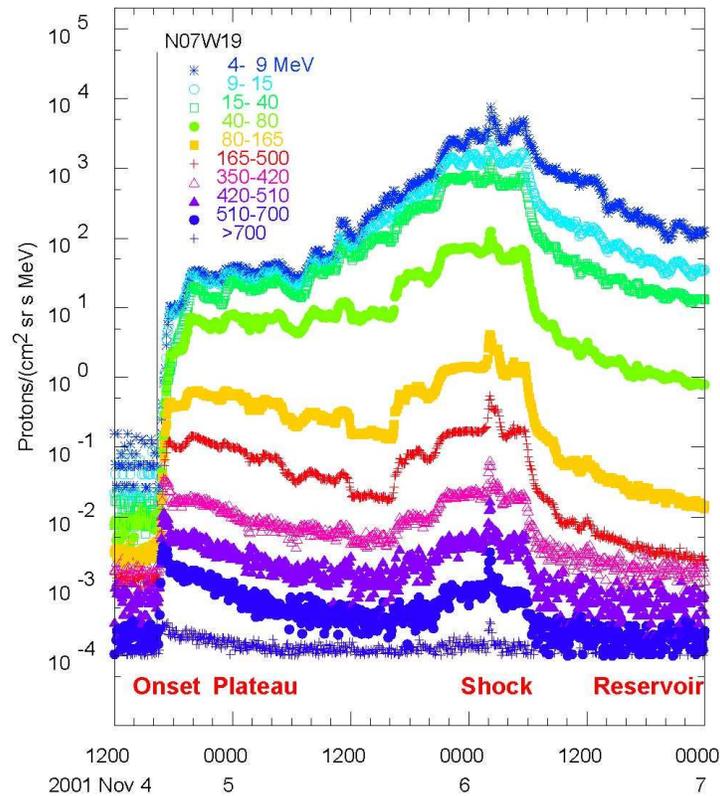

*Figure 5.1.* The four phases of a large SEP event (and GLE) are indicated along the bottom of the figure showing energetic proton intensities in the 2001 November 4 event seen by the *NOAA/GOES* spacecraft.

### 5.1 SEP EVENT ONSETS AND GLES

SEP ions are controlled by "velocity dispersion", *i.e.* ions with the highest velocity arrive first and subsequent particles arrive in inverse velocity order. The left panel in Figure 5.2 shows sequential arrival of $^4$He in various energy intervals at the *Wind* spacecraft in the GLE of 1998 May 6. In the right panel are plotted the onset times of each energy interval for several ion species measured on several spacecraft as a function of the inverse velocity, $v^{-1}$ in min AU$^{-1}$. If we assume that onset time = distance $\times$ $v^{-1}$, the slope of the fitted line represents the path length along the magnetic field from the source, and the intercept gives the solar particle release (SPR) time at the Sun. Using data from a variety of spacecraft it has been possible to analyze ion onset times for 30 GLEs in this way (Reames 2009a, b).

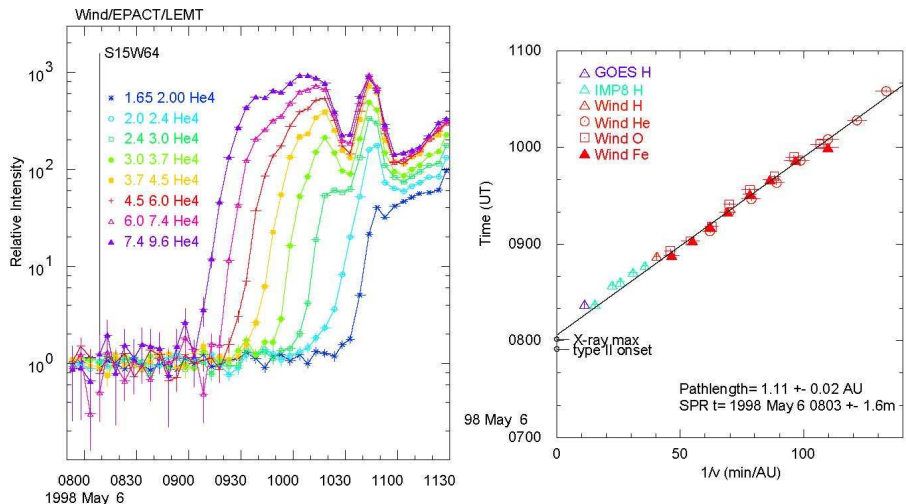

*Figure 5.2.* The left panel shows the arrival of ions of different energy at *Wind* in the 1998 May 6 GLE. Clearly, from its rise time, a "gradual" SEP event onset is *not* very gradual. The right panel shows the onset time of each energy interval *vs.* $v^{-1}$. For the fitted line, the slope is the pathlength and the intercept in the solar particle release (SPR) time at the Sun (Reames 2009a).

Of course this analysis assumes that the earliest arriving particles are unscattered and are focused by the diverging magnetic field to travel at ~0$^{\text{o}}$ pitch angle. With the extremely high intensities in GLEs only a tiny fraction of particles need meet these criteria to provide reliable onset times (see Sáiz *et al*. 2005, Reames 2009a). Recent calculations using the numerical transport model of Ng, Reames, and Tylka (2003) show that velocity dispersion analysis for ions can produce SPR times that are in error by less than ~3 min (see appendix of Rouillard *et al*. 2012). Velocity dispersion analysis uses data that is much more accurate than neutron monitor data alone, which barely rise above the background by <10% in most GLEs. Also, it can be performed in both impulsive and gradual events. Figure 5.3 (Tylka *et al*. 2003) compares the SPR times with solar X-ray and γ-ray data. The left panels compare SPR times (red) for two impulsive events with hard X-ray data (blue) and the right panels compare the SPR times for two GLEs with 4-7 MeV γ-ray data (blue). For the impulsive events, the SPR times agree well with the hard X-ray peaks (there is no γ-ray emission in these two events). However, for the two GLEs, the SPR times fall well after the γ-ray peaks.

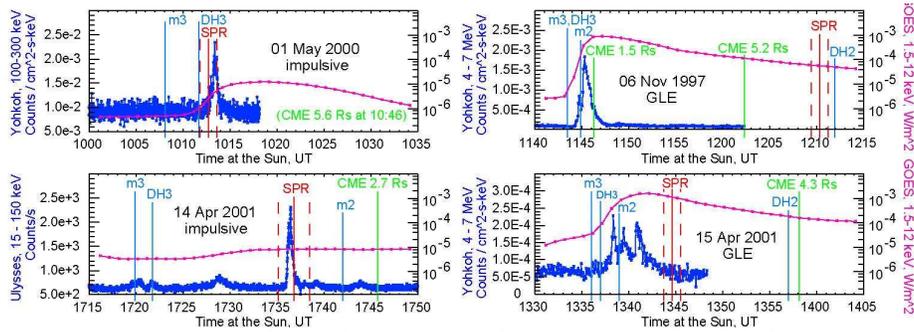

*Figure 5.3*. A comparison of onset timing is shown for two impulsive events (left) and two GLEs (right). The blue curves show intensities of hard X-rays in the left panels and of 4-7 MeV γ-rays in the right panels. Red curves show GOES 1.5-12 keV soft X-ray intensities. Red time lines labeled SPR show the SPR times of the particles, with errors. Other time lines show onsets of type II and type III radio emission and CME observations (after Tylka *et al*. 2003)

We note that non-relativistic electrons may also be used in velocity-dispersion analysis. However, high-energy electron propagation is usually not scatter free (Tan *et al*. 2011), as discussed in Section 4.1, and should not be used.

These results from SEP velocity dispersion analysis strongly confirm the earlier conclusions from the timing of the ~GeV proton release by Cliver *et al*. (1982) on the poor correlation of SEP onset times in GLEs with photon emission from the associated flares. In contrast, for impulsive events, flare and SEP onset times correlate extremely well as seen in the left panels in Figure 5.3 (see also Reames, von Rosenvinge, and Lin, 1985, and Reames and Stone, 1986).

Suppose we divide the GLEs into "late" events where the SPR times are long after the flare and "early" events where any delay is short. In the "late" events the SEPs must surely be shock accelerated and, surprisingly, we receive *no significant particles from the flare* since the SPR time refers to the *first* particles that arrive. The flare seems to contribute nothing. In contrast, for the "early" events, a flare source is possible from the timing. However, timing alone does not determine the source. These "early" events usually also have associated fast CMEs and shock waves (*e.g.* Gopalswamy *et al*. 2012) which must accelerate just as efficiently here as they do in the "late" events. Furthermore, if flares do not contribute *any* SEPs to the "late" events, why would they suddenly contribute in the "early" events? Hence we find no compelling evidence for any flare-accelerated SEPs at 1 AU in any GLE events.

In fact, there is much stronger evidence for the origin of the particles in GLEs. In most SEP studies, GLEs are included along with other large gradual SEP events. Thus, for example, (1) GLEs were included in the early studies of abundances (Meyer 1985) and there are 7 GLEs among the 43 SEP events contributing to the SEP abundances of Reames (1995a) that show the FIP-dependence in the lower panel of Figure 2.1. (2) Eight GLEs are included in the list of 48 *gradual* events in the Reames and Ng study of Z>50 ions. The abundances in these 8 GLEs are *not* enhanced. (3) There are 3 GLEs, all among the electron-poor population, in the study of the e/p ratio by Cliver and Ling (2007). (4) There are 3 GLEs in the study of SEP intensity *vs*. CME speed by Kahler (2001) seen in Figure 1.2 (some events at *Helios* might also qualify). GLEs have very high CME speeds (*e.g.* Gopalswamy 2012), hence high SEP intensities. (5) Three of the five events in the study showing that acceleration began outside 5 solar radii (Kahler 1994) were GLEs, including the huge GLEs of 1989 September 29 and October 24. (6) Shock spectra have a high-energy break downward or knee (Ellison and Ramaty 1985; Lee 2005); spectral breaks have been measured recently in 16 GLEs by Mewaldt *et al.* (2012). (7) There are 13 GLEs among the 44 events in the Tylka *et al*. (2005) study of shock geometry and seed population. (8) A spatially extended (30$^o$) γ-ray source associated with the 1989 September 29 GLE (Vestrand and Forrest 1993) might easily be explained in our modern view by shock-accelerated particles trapped in a reservoir behind the expanding shock that leak back to the Sun. (9) Also, Long-duration γ-ray events (Ryan 2000) might have a similar explanation. Four of the five western-hemisphere sources among these long-duration events were GLEs (Cliver 2006). These γ rays may be giving us a view of shock acceleration from its downstream, solar side. More studies of γ-ray timing and spatial distributions could be extremely helpful in studies of SEP events.

Viewed from the perspective of shock acceleration, the delay of the SPR time after the onset of the type II radio emission is always positive and can vary from a few minutes to as much as ~40 min (Reames 2009b). Using the shock or CME speed, this can be interpreted as the height or radius at which SEP acceleration begins. For the largest GLEs, near central meridian, acceleration begins near ~2 solar radii. At these lower altitudes, higher magnetic fields (Zank, Rice, and Wu 2000) and higher seed-particle densities (Ng and Reames 2008) may contribute to the higher-energy particles. To some extent, higher shock speeds may compensate for the reduced field and density at higher altitudes (Reames 2009b). Separately, Cliver, Kahler, and Reames (2004) found that (non-GLE) SEP events were most strongly correlated with 1-14 MHz type II radio emission corresponding to acceleration above ~3 solar radii. We should also note that a shock is formed and acceleration can really begin when the speed of the CME exceeds the local Alfvén speed. The Alfvén speed (and fast-mode speed) decreases rapidly with radial distance to a minimum of ~200 km s$^{-1}$ at ~ 1.5 solar radii (~100 MHz), rises to a maximum of ~500–700 km s$^{-1}$ at ~3 solar radii (~14 MHz) and decreases thereafter (Mann *et al*. 1999, 2003; Gopalswamy *et al*. 2001; Vršnak *et al.* 2002)

On the flanks of the shock, acceleration times are later and altitudes are higher than at central meridian of the CME, following a broad, roughly parabolic pattern. This is observed both from the distribution of individual events and from the multi-spacecraft observation of a single event at 4 distinct longitudes (Reames and Lal 2010). Delays in the SPR times on the flanks of the shock could be caused by either or both of the following: 1) reduced shock speed on the flanks when a single shock or CME speed was assumed or 2) the shock actually expands to strike field lines at higher altitudes on the flanks, as, for example, shown by the STEREO simulations in the left panels of Figure 3.4 (Rouillard *et al*. 2011, 2012).

Shock waves also accelerate electrons, of course (Kahler 2007), and onsets of the type II and type III bursts differ by only a few minutes in large GLEs (Gopalswamy *et al.* 2005, 2012). Recently, Tan *et al.* (2012b) have shown that the path lengths followed by electrons from GLEs are most consistent with those derived from ions if one assumes that the electrons are accelerated at the time of onset of the type II bursts rather than that of the type III bursts.

## 5.2 PLATEAUS AND THE STREAMING LIMIT

It has been observed for many years that intensities of protons at several MeV seen early in the large gradual SEP events are bounded at an upper limit of ~200-400 protons (cm$^2$ sr s MeV)$^{-1}$ (Reames 1990, Reames and Ng 1998). While this limit applies to events from any solar longitude, it is most obvious for events from central meridian that form an intensity plateau as in Figure 5.1.

Protons streaming out from a shock wave near the Sun generate Alfvén waves that scatter particles coming behind (Stix 1992). Increasing the source proton intensity increases wave growth and the added scattering causes the 1 AU intensity to increase less rapidly. Eventually, with increasing source intensity, the intensity near 1 AU will stop increasing at the "streaming limit." Such a limit is inherent in all of the analytic *equilibrium* shock theories involving wave amplification (*e.g.* Bell 1978; Lee 1983; Sandroos and Vainio 2009) and has been studied in detail by Lee (2005). The streaming limit has also been simulated by numerical transport models that include wave amplification and the full evolution of the pitch-angle distribution (Ng and Reames 1994; Ng, Reames, and Tylka 2003, 2012).

Recent studies of particle spectra on the plateaus of the largest SEP events have shown energy spectra of the ions that are not power-law in form but peak at ~10 MeV amu$^{-1}$ and roll downward toward lower energies, as shown for H and O in the left panel of Figure 5.4. The right panel compares the plateau proton spectrum for the smaller 1998 May 2 GLE with that of the large 2003 October 28 GLE. We can understand these spectra as follows: The wave number of resonant waves is $k \approx B/P\mu$ where $P=pc/Qe$ is the rigidity for a particle of momentum $p$ and charge state $Q$, and $\mu$ is its pitch-angle cosine with respect to the magnetic field $\boldsymbol{B}$. Thus, for example, protons of energy ~10 MeV first stream out at $\mu$~1 and generate resonant waves which scatter subsequent ~10 MeV protons toward smaller $\mu$, where they generate waves at higher $k$ that can resonantly scatter ~1 MeV protons with $\mu$~1 which are just beginning to arrive. Thus high intensities of protons, with ~10 MeV and above, suppress the intensities of ~1 MeV amu$^{-1}$ ions as seen for the events in the left panel of Figure 5.4. When the intensities of ~10 MeV protons are a factor of ~100 lower, as in the 1998 May 2 event, seen in the right panel of Figure 5.4, the wave production is inadequate to suppress the spectrum at low energies. These spectra are well simulated by theoretical models that follow the evolution of the wave amplification and the pitch-angle distributions of the ions (*e.g.* Ng, Reames, and Tylka 2003, 2012).

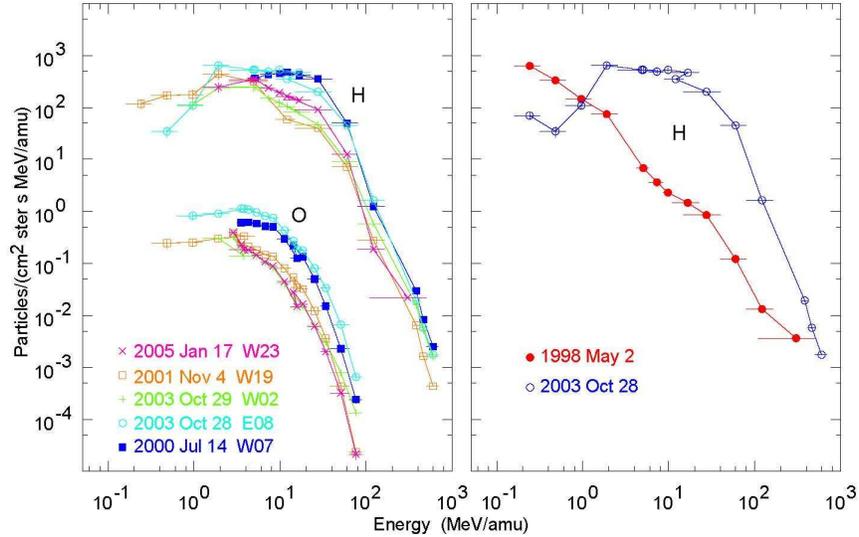

*Figure 5.4.* The left panel compares plateau spectra of H and O in 5 large GLEs. The right panel compares plateau proton spectral shapes for 2 GLEs with large differences in proton intensities at ~10 MeV. High intensities of streaming protons at ~10 MeV produce waves that suppress the spectra at ~1 MeV, low intensities at ~10 MeV do not (Reames and Ng 2010).

Note that the streaming limit is actually a particle transport effect and its value is a function of distance between the source and the observer. When the shock itself approaches the observer, intensities rise toward a peak at the nose of the shock whose intensity is not similarly bounded. An equilibrium streaming limit exists at all energies and the high-energy behavior has important practical value. Lee (2005) finds that the limit varies as $E^{-1}$ and the simpler theory of Sandroos and Vainio (2009) gives $p^{-1}$. Reames and Ng (1998) tried to define a high-energy limit from long-term observations and also found a radial dependence of the intensity consistent with $\sim R^{-3}$. In general, the coupling of the intensities at different energies complicates the process, *i.e.* the intensity at one energy, can depend upon the intensities at all higher energies, as we have seen. Furthermore, the streaming limit we calculate is an equilibrium value and a minimum number of protons must flow past to generate enough waves before this value is established. The intensity of these pre-equilibrium protons may be arbitrarily high if they arrive in an arbitrarily short time. Finally, trapping might allow intensities to rise above the limit, as pointed out by Lario, Aran, and Decker (2008, 2009).

In general, an intensity plateau may be produced by an improving magnetic connection to a weakening shock, commonly near central meridian. The streaming limit only limits the maximum intensity that the plateau can have. The streaming limit applies at any longitude, but it is most clearly observed on the plateau.

5.3 SHOCK WAVES

Shock acceleration is a very old subject that has been reviewed extensively by Jones and Ellison (1991), for example. The theory of interplanetary shocks has been reviewed recently by Lee (2005) and by Zank, Li, and Verkhoglyadova (2007). The physical process of wave amplification by streaming protons (*e.g.* Stix 1992, Melrose 1980) underlies diffusive shock acceleration. The theory of Bell (1978) was adapted to acceleration at interplanetary shocks by Lee (1983), providing a basis for studies of acceleration of protons to ~GeV energies near the Sun (Zank, Rice, and Wu 2000; Lee 2005; Sandroos and Vainio 2007; Vainio and Laitinen 2007, 2008; Ng and Reames 2008; Battarbee, Laitinen and Vainio 2011).

5.3.1 *Modeling Acceleration near the Sun*

Recently, Ng and Reames (2008) calculated the time dependent evolution of SEP acceleration by a quasi-parallel shock wave near the Sun. Evolution of proton and wave spectra at a fast parallel shock is shown in Figure 5.5. This model, based on an earlier SEP transport model (Ng, Reames, and Tylka 2003), is self-consistent in that streaming protons amplify Alfvén waves that subsequently scatter the protons and reduce the streaming. Proton spectra and angular distributions are followed in radial distance and time. Evolution of the proton pitch angle plays an important role in the acceleration. Protons first attain each new, higher energy at a pitch angle near ~90° ($\mu$~0) where they resonate with waves that were amplified by lower-energy protons at smaller pitch angles (*e.g.* $\mu$~1). Thus, lower-energy protons prepare the scattering environment necessary for their subsequent rise to higher energy, an important insight from this model (Ng and Reames 2008).

Quasi-perpendicular shocks are expected to be more efficient than quasi-parallel shocks because the particles can gain additional energy on each traversal by drifting in the $V_s \times B$ electric field of the shock (*e.g.* Decker 1981). If we assume that the knee energies seen in the left panel of Figure 5.5 are shifted by the sec $\theta_{Bn}$ dependence in Equation 2.1, then the knee energy might reach hundreds of MeV at much earlier times for an oblique or quasi-perpendicular shock. However, waves generated upstream will also be more rapidly overtaken by the shock, a possible compensating effect that must be considered numerically.

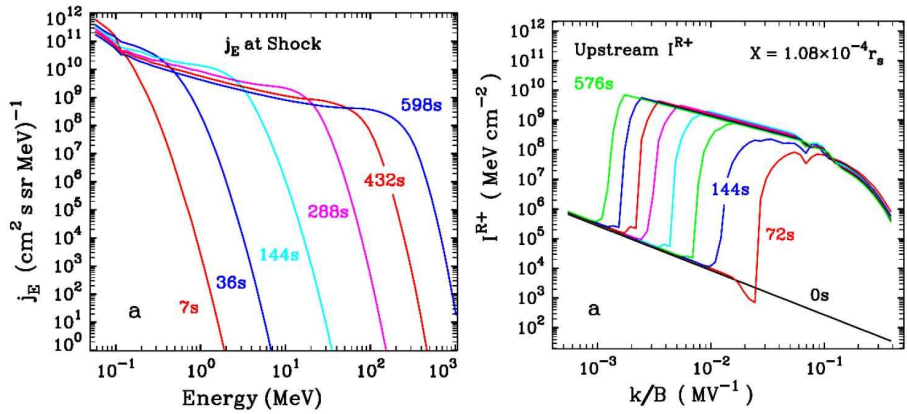

*Figure 5.5* The left panel shows the time evolution of the proton spectrum at a quasi-parallel shock while the right panel shows the corresponding evolution of right-handed outbound Alfvén wave spectrum just upstream of the shock (Ng and Reames 2008).

### 5.3.2 *Interplanetary Shocks*

Traveling interplanetary shock waves near Earth are the local continuation of the CME-driven shock waves that produce gradual SEP events. These shocks provide an opportunity to directly measure, *in situ*, the properties of accelerated particles together with the characteristics of the shock that accelerated them under an extremely wide variety of conditions (see *e.g.* Berdichevsky *et al.* 2000). Gosling *et al.* (1981) first showed that the low-energy spectrum of accelerated particles forms a continuum with the spectrum of the seed population in the solar wind, from which it is primarily derived. Desai *et al.* (2003) showed that low-energy ion abundances near the shock peak were much more closely correlated with ambient abundances upstream of the shock than with the abundance of the corresponding elements in the solar wind, as expected from our discussion of the seed population in Section 2. Desai *et al.* (2004) found that the energy spectra at the shocks were better correlated with those upstream than with those expected from the shock compression ratio.

The choice of a location to measure the ambient, or background, abundances and spectra is always difficult. If it is chosen prior to the time when the shock leaves the Sun, perhaps ~2-3 days prior to the shock arrival time, then solar rotation insures that background is sampled at a longitude of 26-40 degrees to the west of the longitude of the shock peak sample. If it is chosen hours prior to the shock arrival, background will include particles accelerated earlier by the same shock. Neither assumption is ideal; either causes variations in the results.

In effect, the re-acceleration of ions from the seed population found in the reservoir of an earlier event evokes the classical two-shock problem considered, for example, in the review by Axford (1981) and more recently by Melrose and Pope (1993). Here, the equilibrium distribution function $f(p)$ of momentum $p$ of accelerated particles downstream of the shock with compression ratio $r$ is

$$f_a(p) = ap^{-a} \int_0^p dq \; q^{a-1} \phi(q) \tag{5.1}$$

where $a=3r/(r-1)$ and $\Phi(p)$ is the injected distribution. If we take $\Phi(p)$ as a delta function at $p_0$ we find a power law spectrum $f_a(p) \sim (p/p_0)^{-a}$ after the first shock. If we reapply Equation 5.1 injecting $f_a(p)$ into a shock with compression ratio $r'$ and let $b=3r'/(r'-1)$, we find that integrating the power law gives

$$f_{a,b}(p) = \frac{kab}{p_0(b-a)} \left[ \left(\frac{p}{p_0}\right)^{-a} - \left(\frac{p}{p_0}\right)^{-b} \right] \qquad \text{for } a \neq b \tag{5.2}$$

The intensity is $j(E)=p^2 f(p)$.

Note that Equation 5.2 is symmetric in the powers *a* and *b*, and will be dominated by the shape of the hardest, flattest spectrum, either the background (*i.e.* a) or the new shock, b. Thus it is no surprise that one finds local shock spectra that are dominated by the shape of the background spectrum (Desai *et al.* 2004; Reames 2012). A further complication occurs when we include a spectral knee with a factor like exp(-$E/E_{0i}$), with $E_{0i}$ defined in Equation 2.1. At energies above the knee, observers will find spectra that are much steeper than either the background or the expected equilibrium shock spectra.

These possibilities for spectral shapes were considered in the observations of Reames (2012), who studied $^4$He spectra of ~1-10 MeV amu$^{-1}$ in 258 *in situ* interplanetary shocks observed by the *Wind* spacecraft. An interesting feature of this study was determining which shock parameters were important to produce measurable particle acceleration. The left panel in Figure 5.6 shows a histogram of the shock speed distribution for all of the shocks and for the subset that showed measurable particle acceleration. High shock speed was the strongest determinant of acceleration, followed by high shock compression ratio, and large $\theta_{Bn}$. Measurable acceleration was more than twice as likely for shocks with $\theta_{Bn}>60^o$ as for those with $\theta_{Bn}<60^o$. Quasi-parallel shocks, *i.e.* small $\theta_{Bn}$, probably were more likely to have knee energies below the energy interval of observation. Recently, Zank *et al.* (2006) have suggested that "higher proton energies are achieved at quasi-parallel rather than highly perpendicular interplanetary shocks within 1 AU." The recent *in situ* observations (Reames 2012) seem to show the opposite; quasi-perpendicular shocks are favored.

The right panel in Figure 5.6 shows the peak shock intensity of 1.6-2.0 MeV

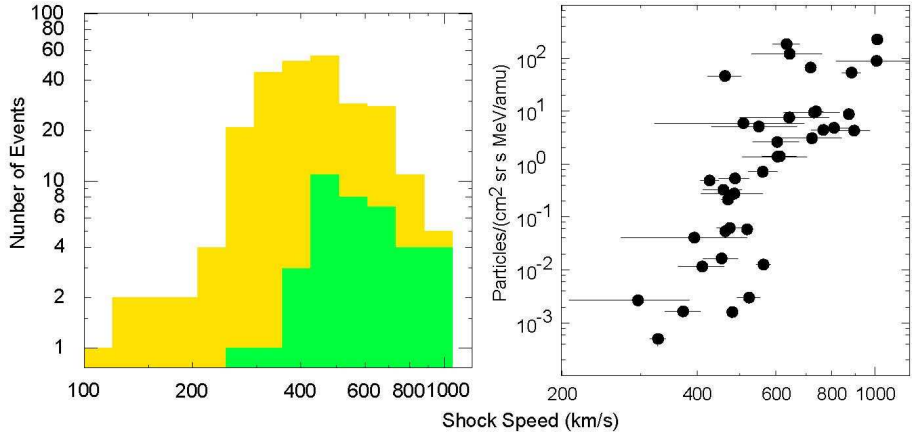

*Figure 5.6.* The left panel shows the distribution of shock waves at 1 AU with measurable intensities of >1 MeV amu$^{-1}$ $^4$He *vs.* shock speed (green) within the total distribution of 258 shock waves *vs.* shock speed (yellow and green) observed by the *Wind* spacecraft. The right panel shows the background-corrected peak intensity of 1.6-2.0 MeV amu$^{-1}$ $^4$He *vs.* shock speed for *in situ* shocks. Shock speed is the strongest determinant of accelerated intensity for local shocks; this mirrors the correlated behavior of peak intensity *vs.* CME speed in Figure 1.2.

amu$^{-1}$ $^4$He as a function of shock speed. The shock speed has a correlation coefficient of 0.80 with intensity. This correlation for *in situ* shocks mirrors the correlation of peak proton intensity with CME speed in Figure 1.2 as modified by Rouillard *et al.*(2012) and shown in the lower right-hand panel of Figure 3.4.

Intensities of accelerated particles are not well predicted by acceleration theory. The rate of injection of seed particles is generally treated as an adjustable parameter – more input results in more output. However, streaming protons and increasing wave intensities must be bounded. At a few powerful shock waves, such as 1989 October 20, it has been observed that the energy in energetic particles exceeds that in the plasma and magnetic field (Lario and Decker 2002). Those authors have suggested that the peak intensities of particles up to 500 MeV are simply trapped in a region of low density and field near a shock. But, how did they get there? I would suggest that the particles are in the process of destroying (*i.e.* pushing apart *B* at) the shock that accelerated them. Another shock where the particle energy exceeds the magnetic energy is that of 2001 November 6, in Figure 5.1 (C. K. Ng, private communication), where the sharp particle peak shows a shock that is still clearly intact. This is the issue of "cosmic-ray-mediated" shocks discussed by Terasawa *et al.* (2006) for two additional interplanetary shocks. This is a fascinating process that can be best observed, *in situ*, at interplanetary shocks.

A recent paper by Desai *et al.* (2012) examined ~80-300 keV amu$^{-1}$ CNO spectra and associated magnetic power spectra at 74 shocks *in situ*. They found that six quasi-parallel shocks ($\theta_{Bn}<70^o$) had strong wave enhancements in the magnetic power spectral densities around the proton gyro-frequency. Four quasi-perpendicular shocks ($\theta_{Bn}>70^o$) had no spectral enhancements and slight or no spectral hardening that the authors associated with shock-drift acceleration. The remaining 64 shocks "exhibit mixed particle and field signatures."

## 5.4 RESERVOIRS

The concept of a reservoir has been important for many aspects of SEP events. Most profoundly, it has altered our thinking so that the idea of a slow time decay of an expanding reservoir has largely replaced the idea of slow parallel diffusion of the particles through preexisting waves. *Reservoirs make gradual SEP event decays gradual*. However, reservoir intensity uniformity (usually within a factor of ~2 or so) has also forced us to consider slow cross-field transport.

Spectral invariance suggests that SEPs do not leak from a typical reservoir significantly. If high-energy particles, which might encounter a boundary more often, were able to leak out, the spectrum would soon steepen, contrary to the observation of temporal and spatial shape invariance. Thus any leakage must be slow for the energy region below ~100 MeV that has been studied. Well over half of large gradual SEP events exhibit fairly strict spectral invariance. About half of the remaining events do show slow spectral steepening with time that can come from (1) continuing acceleration by a weakened shock, (2) preferential leakage of high-energy particles, or (3) slower cross-field transport of lower-energy particles.

In Figure 5.7 we show the evolution toward spectral invariance in five large SEP events from different solar longitudes. The lower panels in the figure show the intensity of protons in 8 energy channels while the upper panels show the ratio of intensities of each of the 7 lower-energy channels divided by that of the highest-energy (43-63 MeV) channel. When the ratios become horizontal, that part of the spectrum is invariant in time. Note that in some of the events the higher-energy portion of the spectrum attains invariance while the lower energies continue to evolve for the reasons enumerated above.

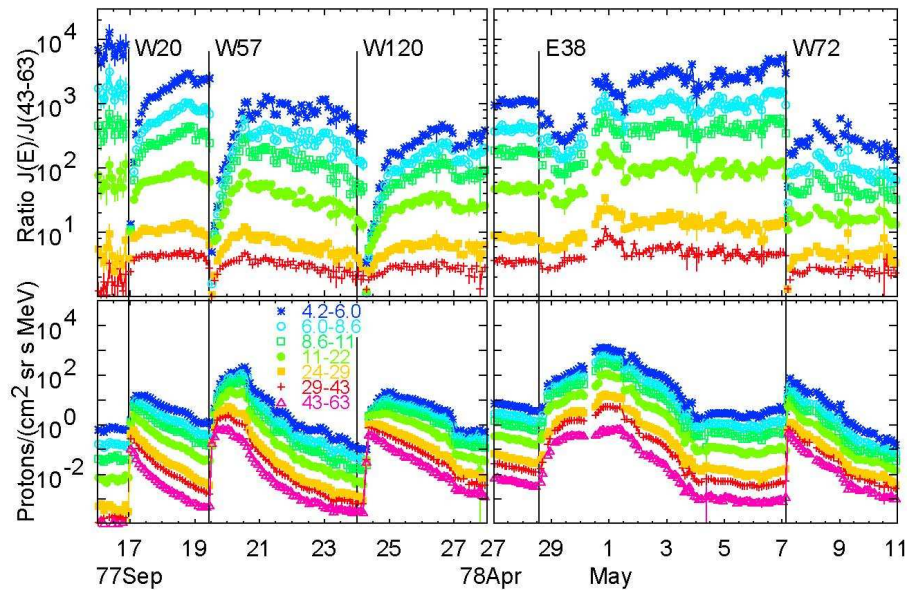

*Figure 5.7.* The lower panels show intensities in 8 energy channels from IMP 8 during 5 large SEP events. Vertical lines mark event onsets at the indicated solar longitude. The upper panels show the ratio of the intensities of the 7 lowest-energy channels divided by the intensity of the 43-63 MeV channel. Ratios become horizontal during periods of spectral invariance. Closely spaced ratios indicate a hard spectrum.

Magnetic trapping by CMEs must apply to SEPs from both impulsive and gradual events. Once a magnetic bottle is formed, energetic particles from nearly any sources can fill it, even ACRs, given a sufficiently long time. However, gradual events have several advantages in producing observable reservoirs, such as: (1) much higher SEP intensities, (2) continuing acceleration, (3) broad initial spatial distribution, and (4) a self-contained CME. SEPs from an impulsive event must be "accidentally" trapped by a preceding CME to form a reservoir, but this is not terribly unlikely. At solar maximum an average rate of 2.5 CMEs day$^{-1}$ emitted from the Sun (Webb and Howard 1994) would produce ~1 CME AU$^{-1}$ in every direction if emitted randomly, and if each occupied ~1 steradian (Reames 1999). Such accidental preceding CMEs *upstream* of the accelerating shock (*e.g.* beyond 1 AU) may also trap SEPs in a gradual event, especially upstream of the eastern flank of the shock, which otherwise appears to be open (see Figure 3.1).

The onset of the invariant spectrum upstream of the shock in the left panel of Figure 3.2 occurs at a tangential field discontinuity, *i.e.* the intersection of two different magnetic regimes, not at a compression region. However, with multiple ejecta upstream it is not too surprising that leakage from reservoirs is minimal since the SEPs would have to run a gauntlet of earlier CMEs with magnetic compression regions to get far from the Sun.

A reservoir formed by one event (either impulsive or gradual) can contribute the seed population for acceleration by the shock of a second event as discussed in Section 5.3.2. This causes a complex replication of the abundances. At a given energy, the intensity of each species is formed by integration over lower-energy seeds (Equation 5.1). This may also cause a replication of the spectral index, when the spectrum of the first event is harder than that which could be produced by the second shock. An example was shown by Reames, Ng, and Tylka (2012) for which the high-energy proton spectrum early in the GLE of 1977 September 24 resembles that seen in the reservoir behind the large GLE that began on September 19. As they suggest, it is possible for an SEP event arising from a shock wave with a modest compression ratio to be promoted to become an "accidental GLE" simply by sampling particles from the reservoir of a previous GLE with a hard spectrum.

Multiple CMEs can interact to form merged interaction regions (*e.g.* Burlaga, *et al.* 2003) when fast CMEs overtake slower ones. These can be large complex regions that modulate the intensities of galactic cosmic rays, but they also efficiently contain SEPs to form reservoirs behind them. In a series of large SEP events, intensities can build up and span an extremely large region of space. Such are the large "super-events" (Mueller-Mellin, Roehrs, and Wibberenz 1986; Dröge, Muller-Mellin, and Cliver 1991). However, super-events are defined by their azimuthal spread, not the successive buildup of intensities in a single reservoir.

Several authors have suggested that particle acceleration is enhanced by the presence of a previous CME or SEP event (Kahler, 2001; Gopalswamy *et al.*, 2002, 2004; Cliver, 2006; Li *et al.* 2012). The presence of a reservoir full of SEPs from the previous event must surely be a factor. Integration over this distribution, as specified by Equation 5.1, can provide a significant enhancement and might even provide a harder spectrum than would be expected from the compression ratio of the second shock. Furthermore, *the newly accelerated particles are trapped by the same reservoir that provided the seed particles*, further increasing the intensity. In the GLE of 1979 August 21 (Cliver 2006), SEP intensities from an energetic event ~3 days earlier reach a peak at the time of passage of a shock and CME on August 20 then form an invariant-spectral reservoir with a hard proton spectrum. After the onset of the W40 GLE on August 21, the spectrum hardens further, with proton intensities increasing by an order of magnitude above ~100 MeV, but only a factor of <2 below 10 MeV.

## 6. Radiation Hazard

High-energy SEPs from large gradual events are of more than academic interest to space travel since they can constitute a significant radiation hazard for astronauts and equipment, especially beyond the Earth's magnetic field (see reviews Barth, Dyer, and Stassinopoulos 2003; Xapsos et al. 2007; Cucinottta *et al.* 2010). Furthermore, fragmentation induced by high-energy protons in the upper atmosphere can produce penetrating radiation, especially neutrons, which threaten the passengers and crew of high-altitude aircraft on trans-polar routes. Protons of ~150 MeV can penetrate 20 gm cm$^{-2}$ (7.4 cm) of Al or 15.5 cm of water (or human flesh). Such protons are considered to be "hard" radiation, in that they are very difficult to shield, and they are orders of magnitude more intense than the GeV protons that define a GLE. Essentially the entire radiation risk to humans in space from SEP events comes from protons in the energy region above about ~50 MeV, or "soft" radiation. This is where protons begin to penetrate spacesuits and the skin of spacecraft.

The location of the energy spectral knee is the most important single factor in the radiation dose and in the depth of penetration of the ions. Figure 6.1 compares proton spectra for two large SEP events. The hazardous portion of the spectrum for the 1998 April 20 event is shaded yellow. The region of *additional* dose in the 1989 September 29 event is shaded red. Note that both events have similar intensities below ~60 MeV; in fact, the integral fluence above 10 MeV is actually larger in the 1998 April event. Even behind 10 g cm$^{-2}$ of material astronauts would receive a dose of ~4 rem hr$^{-1}$ (40 mSv hr$^{-1}$) at intensities in the 1989 September event.  The *annual* dose limit for a radiation worker in the United States is 5 rem (50 mSv) (see review Cucinotta *et al.* 2010).  There are several events of this intensity in each solar cycle.  *Knowledge of spectral knee energies is important.*

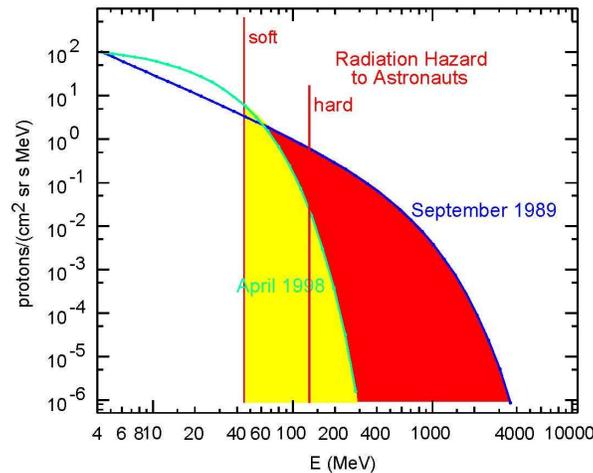

*Figure 6.1* Spectra of the SEP events of 1998 April (green; Tylka *et al.* 2000) and 1989 September (blue; Lovell, Duldig, and Humble 1998) are compared. Typical energies of "soft" and "hard" radiation are shown. The hazardous portion of the spectrum of the April event is shaded yellow and the *additional* hazardous radiation from the September event is shaded red.

In addition to the peak intensity, the duration of the events, determined by the solar longitude and by trapping in a reservoir, is a factor in the radiation dose.  Tan, Reames, and Ng (2008) compared the time profiles of protons up to ~500 MeV for events that did and did not show evidence of particle reflection outside 1 AU.  Differences were seen for the otherwise similar events of 2002 April and August for which Fe/C ratios are shown in the left panel of Figure 2.2.

The study of the probability of large SEP events appeared to be greatly enhanced by the observation of nitrates, thought to be generated by SEP events, captured in polar ice over the last ~400 years (McCracken *et al.* 2001a, b).  Unfortunately, recent studies by Wolff *et al.* (2012) surrounding the Carrington (1860) event have shown the SEP association with nitrates to be in error; the nitrate peaks are probably caused by biomass burning plumes.  Under these circumstances we must be content with the more limited statistics of recent SEP studies (*e.g.* Reedy 1996, 2006, 2012; Schrijver *et al.* 2012).

One way to show the probability of high intensities is by binning hourly-averaged data over a solar cycle in intensity bins as shown in Figure 6.2.  Each panel shows the number of hours in each intensity interval for protons of the energy interval shown.  Above the streaming limit, one sees intensities only at a few shock peaks, so there is a significant drop in occurrence rate. Strong shock peaks are only seen over a limited interval of solar longitude. The dashed black lines in Figure 6.2 are fits to data below the streaming limit that decrease as the ~0.4 power of the intensity. This effect is often expressed as a rate of change in the number for a given change in the intensity, which decreases as the 1.4 power of the intensity in this case. Cliver *et al.* (2012) have recently compared these distributions for different size measures of SEP events and of hard and soft solar X-ray events.

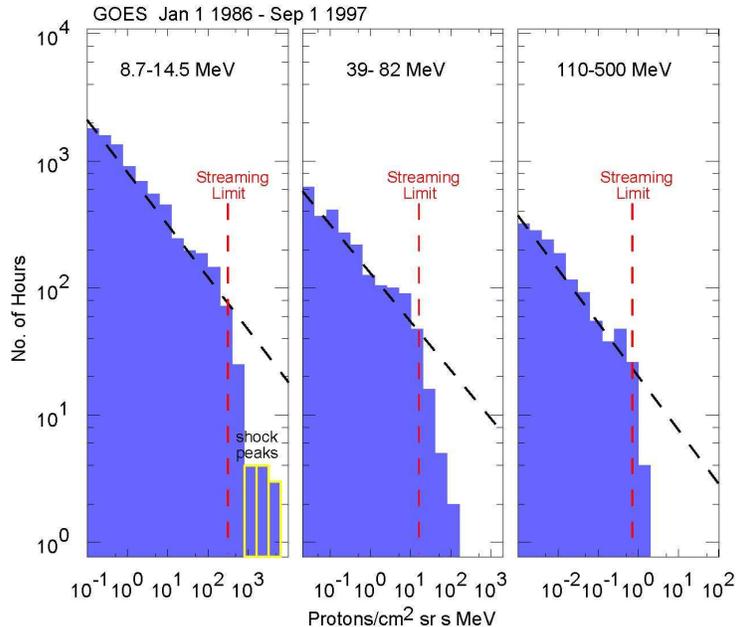

*Figure 6.2* For each of three energy intervals, panels show the number of hours spent at each intensity level during an ~11 year period.  Intensities of the streaming limit are indicated for each energy bin (Reames, Tylka, and Ng 2001).

There is little hope for a useful and *definitive* warning at the onset of an SEP event. We have seen in Section 5.1 that a flare sometimes precedes the SPR time by up to ~30 min, but this delay is unreliable and, at the present time, we have no good way to predict the intensity of ~150 MeV protons from flare observations. It has been suggested that relativistic electrons be used as a signal (e.g. Posner *et al.* 2009). This might be an early warning for slow ~1 MeV protons that take hours to arrive, but ~150 MeV protons arrive within 8 minutes of the relativistic electrons or protons. Since relativistic electrons may be delayed by strong scattering (Tan *et al.* 2011), the onset of a type II burst may be a preferable indicator but it still gives only an 8-min warning.

The existence of the *streaming limit* does place a bound on intensities for a time of perhaps *a day*, at least until the shock approaches, but streaming-limited intensities are already rather high. However, this seems to be the best limit we presently have to offer, although the high-energy limit needs further study.

The peak intensities of protons in the interval ~110-500 MeV are correlated with the CME speed with a correlation coefficient of ~0.5. Knowledge of the shock speed on the magnetic field line to Earth might improve this correlation. Thus, an early measure of even the CME speed or of the shock speed from the drift rate of the type II burst would be an indicative forecast of proton intensity. Still, this will only give a warning of ~10 min.

While accurately predicting hazardous SEP events well in advance remains only a goal, we can certainly help by improving our understanding the conditions where, when, how, and why acceleration takes place in exceptional events. Most importantly, we can study the relationship of SEP intensities and spectra with measurable parameters of the shock, plasma, and seed population and their spatial distribution, as well as the association of SEP duration with magnetic trapping by preceding CMEs. Study of these factors for >150 MeV protons has been minimal.

## 7. Conclusion

After 50 years, the two sources of SEP acceleration proposed by Wild, Smerd, and Weiss (1963) remain valid. SEPs associated with type III bursts are certainly dominated by electrons, as suggested, although their truly unique character comes from enhancements in $^3$He and heavy elements and the window that provides on the physics of magnetic reconnection and of resonant wave-particle interactions in impulsive flares and jets. SEPs associated with type II bursts produced by shock waves driven by fast CMEs are indeed proton dominated, though they also tell us about coronal abundances and about the spatial and temporal distribution of shock acceleration and of particle transport in a complex, CME-filled inner heliosphere.

This manuscript is based on a lecture given by the author upon receipt the 2012 George Ellery Hale Prize from the Solar Physics Division of the American Astronomical Society. It was funded, in part, by NASA grant NNX08AQ02G. The author thanks Ed Cliver, Steve Kahler, Frank McDonald, Chee Ng, Lun Tan, and Allan Tylka for their helpful comments and for discussions relating to this manuscript.42